\documentclass{emulateapj}
\usepackage{amsmath}
\usepackage{amstext}
\usepackage{apjfonts}
\usepackage{amssymb}
\usepackage{xspace}
\usepackage{hyperref}
\usepackage{gensymb}
\usepackage{xstring}
\usepackage{natbib}
\usepackage{multirow}
\usepackage{graphicx}
\usepackage{epstopdf}
\usepackage{epsf}
\usepackage{subfigure}
\usepackage[normalem]{ulem}
\usepackage[usenames,dvipsnames]{xcolor}
\usepackage[utf8]{inputenc}
\usepackage[bottom]{footmisc}

\def\H2{\rm H_2}
\def\Msun{{\rm M_{\odot}}}
\def\M1500{M_{1500}}
\def\chimp{h^{-1} {\rm cMpc}}

\begin{document}
\title{The Role of Quasar Radiative Feedback on Galaxy Formation during Cosmic Reionization}
\author{Huanqing Chen\altaffilmark{1,2}}

\altaffiltext{1}{Department of Astronomy \& Astrophysics, The University of Chicago, Chicago, IL 60637, USA; hqchen@uchicago.edu}
\altaffiltext{2}{Kavli Institute for Cosmological Physics, The University of Chicago, Chicago, IL 60637 USA;}

\begin{abstract}
Recent observations have found that many $z\sim 6$ quasar fields lack galaxies. This unexpected lack of galaxies may potentially be explained by quasar radiation feedback. In this paper I present a suite of 3D radiative transfer cosmological simulations of quasar fields. I find that quasar radiation suppresses star formation in low mass galaxies, mainly by photo-dissociating their molecular hydrogen. Photo-heating also plays a role, but only after $\sim$100 Myr. However, galaxies which already have stellar mass above $10^5 \Msun$ when the quasar turns on will not be suppressed significantly. Quasar radiative feedback suppresses the faint end of the galaxy luminosity function (LF) within $1$ pMpc, but to a far lesser degree than the field-to-field variation of the LF. My study also suggests that by using the number of bright galaxies ($\M1500<-16$) around quasars, we can potentially recover the underlying mass overdensity, which allows us to put reliable constraints on quasar environments.
\end{abstract}

\section{Introduction}
Quasars during the epoch of reionization (EoR) are fascinating objects to study. Their absorption spectra encode rich information, such as quasar lifetime, underlying mass distribution, and the ionization state and temperature of the intergalactic medium (IGM) \citep[e.g.,][]{bolton07,bolton12,keating15,kakiichi17,davies18,eilers18}. Studying these z$\sim6$ quasars helps us understand how the first structures are formed in the cosmic history, and how and when the universe underwent the cosmic ``dark age'' to the ionized universe we see today.

 In the past two decades, a race has begun to detect the most distant quasars. Exciting progress has been made: there are more than a hundred of such quasars found \citep[e.g.,][ and the references therein]{fan00,jiang16,banados16,matsuoka18,yang19}, and more follow-up observations are on the way. Apart from studying the property of quasar host galaxies \citep[e.g.,][]{wang13,venemans17,walter18}, there are extensive efforts to map a wider field of the quasar to answer a key question of structure formation: how overdense is the environment of the first quasars?

According to the hierarchical structure formation picture \citep{lacey93} of $\Lambda$CDM cosmology, small structures form first in the highest density peak, followed by larger scale structures collapsing around them. Therefore, luminous quasars, the most prominent small structures, are thought to trace biased overdense regions of the early universe. Consequentially, more galaxies are expected to be detected surrounding quasars \citep{morselli14}. However, observationally it is not always the case. For example, using HST ACS data, \citet{kim09} found three out of five quasar fields show no excess of galaxy candidates, known as $i$-dropout objects.  Studies basing on large field-of-view Subaru Suprime-Cam data also show underdensity of  Lyman-$\alpha$ emitters (LAE) surrounding some quasar fields at $z>6$ \citep[e.g.,][]{ota18,goto17} Recent ALMA observations of 35 quasar fields also show no evidence for millimeter continuum sources within 140 pkpc of the quasars \citep{champagne18}. These findings challenge our basic understanding of how the first compact objects and sub-sequential larger structures are formed in the universe.

Why so many $z\sim6$ quasars do not trace galaxy overdensity? A widely acknowledged explanation is that quasar radiation feedback suppresses galaxy formation  \citep[][]{kashikawa07,goto17} . The idea is simple: quasar radiation can destroy molecules and heat the gas, prevent them from cooling and contracting to form stars. Previously, \citet{kitayama00,kitayama01} have used 1D spherically symmetric simulation to study this problem, and gained the key insights that the collapse time of low mass halos are significantly delayed. \citet{shapiro04} used 2D axisymmetric adaptive mesh refinement (AMR) radiative transfer (RT) code to study how ionization front (I-front) impact minihalos. They found that a minihalo $M_h=1\times 10^7 \Msun$ 1 pMpc away from a quasar will ``photo-evaporate'' \citep{doroshkevich67,couchman86,okamoto08} in $\sim 100$ Myr.
although other studies like \citet{kikuta17} find no sign of feedbackdown to $L_{\rm Ly\alpha} \sim  10^{41.8} ~\rm~erg s^{-1}$

Although these studies revealed the importance of radiative feedback of quasars, they are idealized. For example, real galaxies are not spherical but have disk-like structures, which affect the radiation intensity inside. Cosmological substructures are also potentially important because they may shade halos behind them. Also, “galaxies” in these simulations are hard to be directly compared to real observed galaxies since they did not directly model star formation. So far, there is no consensus about to what degree quasar radiative feedback can impact the observables.  Observationally, there have been efforts to constrain the effect of quasar radiative feedback. \citet{kikuta17} find no sign of feedbackdown to $L_{\rm Ly\alpha} \sim  10^{41.8} ~\rm~erg s^{-1}$. \citet{decarli17, farina17} find rapid star formation galaxies in close projected distance ($\lesssim 50 \rm kpc$) to the quasar host. 

\begin{figure*}

 \centering
    \centering
     {\includegraphics[width=6.6cm]{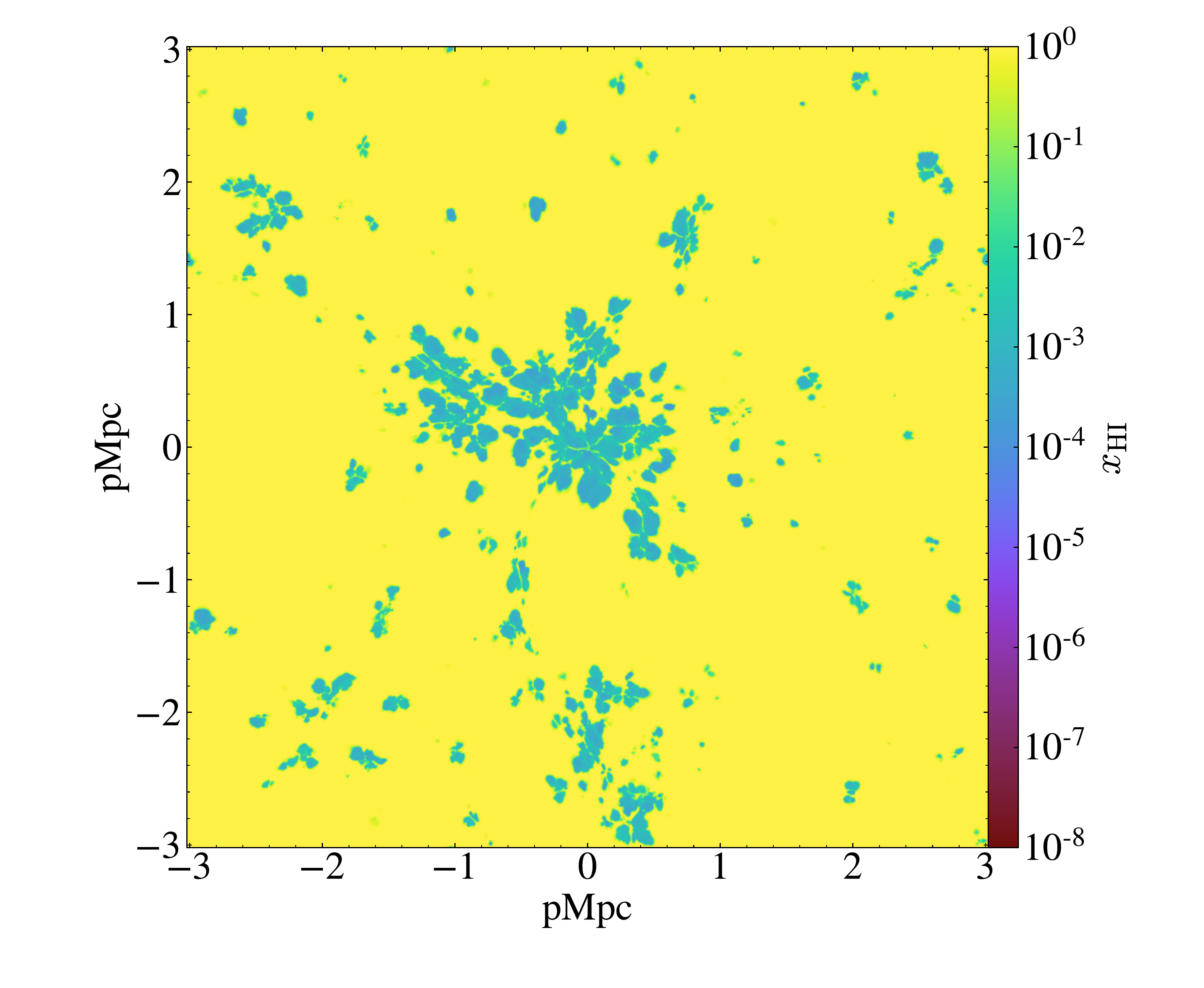}}\hspace{-1.24cm}
     {\includegraphics[width=6.6cm]{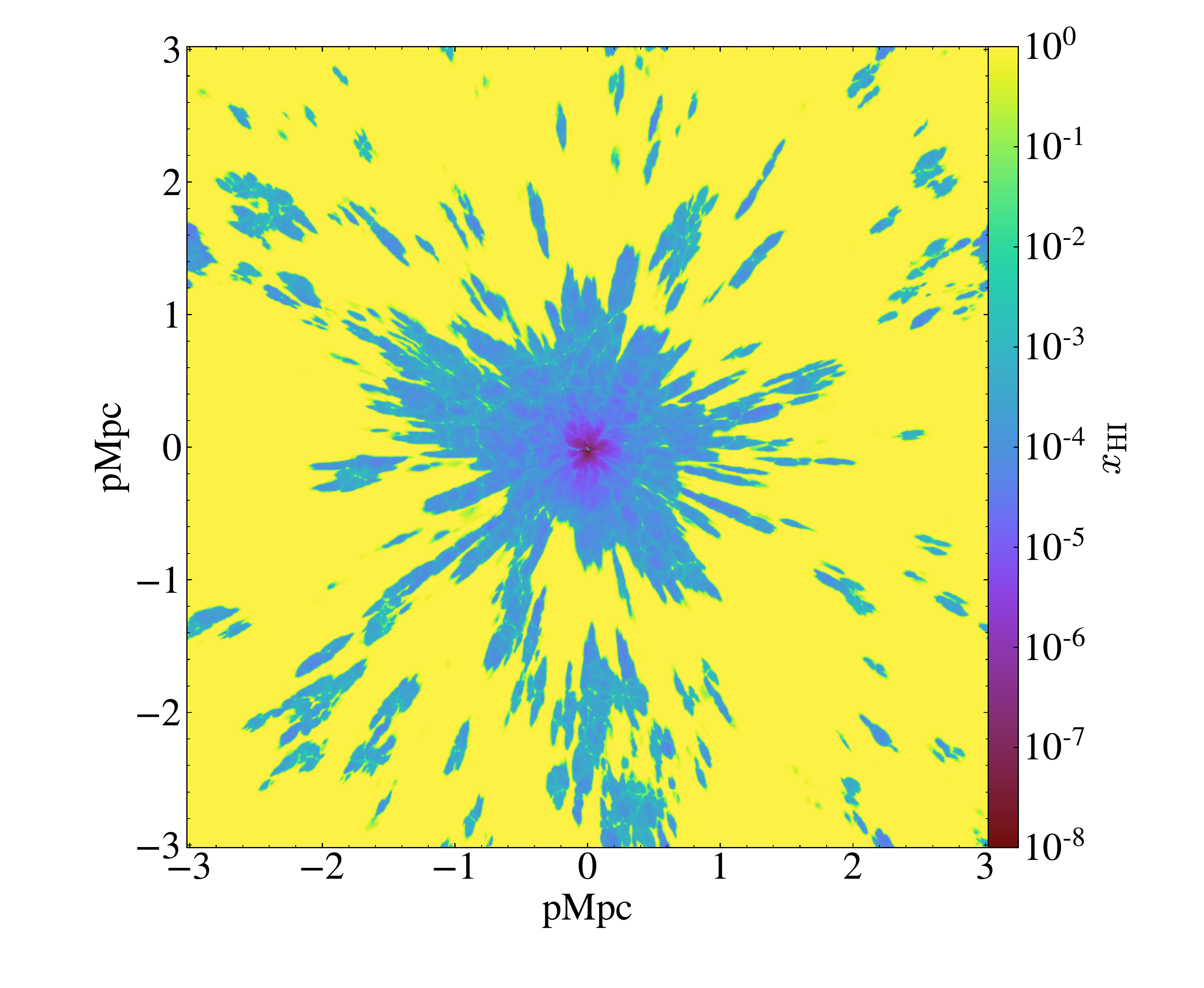}}\hspace{-1.24cm}
     {\includegraphics[width=6.6cm]{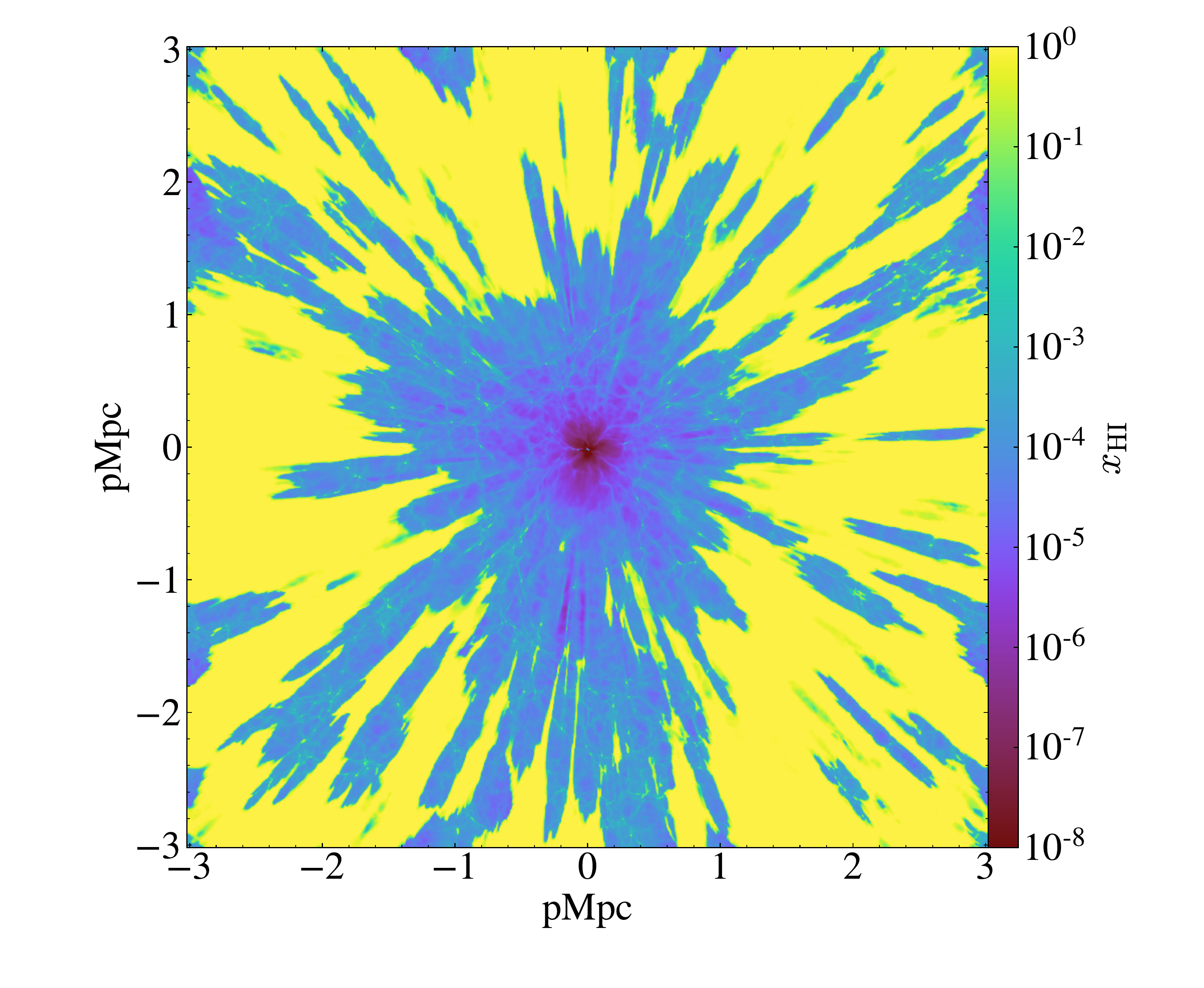}}

\caption{Neutral fraction slice centered on the quasar host at z=8.6 in simulations NoQ (left), Q\_z9\_L(middle) and Q\_z9\_H (right), showing the entire simulation boxes of $6$ physical Mpc (or $40 \ \chimp$). All three panels share the same colorbar as the right panel. At this moment the quasar has been shining for $t_Q=30 \rm Myr$ in Q\_z9\_L and Q\_z9\_H runs. In the left panel, the ionized bubbles are created purely by galaxies, which are strongly clustered around the massive galaxies. In the middle and the right panels, the quasar radiation dominates the quasar surroundings and creates a large ionized region.}\label{fig:xHI}
\end{figure*}

This problem needs to be reevaluated more thoroughly in theory, using more rigorous methods, if we want to understand the true mass overdensity of quasar environments. Equally as important, radiative feedback on galaxy formation itself is a key physical process to study. 
This suppression effect from radiation has previously been overlooked, but recently using ray-tracing radiative transfer simulation, \citet{emerick18} has shown that radiation has a significant effect on star formation in dwarf galaxies. 
\citet{costa19} has found that radiative feedback from stars can ``puff-up'' satellite galaxies, rendering them more vulnerable to tidal stripping.
In quasar proximity, radiation feedback may be even more prominent \citep{proga15}. 

In this paper I present a suite of simulation of galaxy formation in quasar proximity in a cosmological context: I run the full cosmological simulation with one quasar turns on. This work is a follow up of the ``Cosmic Reionization on Computer'' (CROC) project \citep{gnedin14a,gnedin14b}, which use the state-of-the-art hydrodynamic cosmological code with fully coupled RT solver to simulate EoR. I run a suite of simulations with different quasar properties and compare them with the original CROC simulations without quasars. 

This paper is organized as follows: In Section \ref{sec:method} I describe the simulations, including the star formation prescription and quasar model, as well as how I build halo catalogs and calculate galaxy properties. In Section \ref{sec:galaxyProp} I show the galaxy properties in the simulations, including the star formation histories (SFHs) and gas properties. In Section \ref{sec:obs} I show the galaxy luminosity functions in quasar proximity. In Section \ref{sec:discussion} I discuss the sub-grid recipe in the simulation in more details, which helps explain the results, and compare the results to previous studies. A summary and conclusion is offered in Section \ref{sec:conclusion}.

\section{Methodology}\label{sec:method}
\subsection{Simulation Code}
The simulations use Adaptive Refinement Tree (ART) code \citep{kravtsov97,kravtsov02,rudd08}, which uses adaptive mesh to achieve high spatial resolution. The simulations presented here are in a volume of $40 \ \chimp$, which is comparable to the typical size of a quasar proximity zone. 
The number of dark matter particles is $1024^3$, with mass resolution $6.2\times10^6 \Msun$. 
The grid size of the minimum refinement level is 39 ${h^{-1} {\rm ckpc}}$, smaller than the filtering length of the IGM at $z>6$ \citep{gnedin98,gnedin00}, therefore capable of capturing the gas motion of large scale structure formation. 
There are additionally seven levels of adaptive refinement, down to a peak resolution of 100 physical parsecs at $z=6$ \citep{gnedin16}, capable to resolve galaxy disks in the radial direction. Radiative transfer in the simulation is coupled with gas dynamics using Optical Thin Variable Eddington Tensor  \citep[OTVET,][]{gnedin01}, an approximate momentum tensor method which lowers the computation cost.

All the simulations in this study use the same parameters of star formation and stellar feedback as the fiducial CROC simulations, which have shown good agreement with observed galaxy UV luminosity functions at $z=6\sim 10$ and the Gunn-Peterson optical depth at $z<6$. 
I briefly describe the star formation and stellar feedback prescription in CROC and refer to the readers to the methodology paper for more details \citep{gnedin14a}. The star formation model is motivated by the empirical Kennicutt–Schmidt law, where the star formation rate (SFR) is proportional to the surface density of molecular hydrogen \citep[][and references therein]{schmidt59,kennicutt98,colombo18}. The molecular hydrogen is calculated using the fitting formula from \citet{gnedin14c}, which is a function of UV radiation intensity, dust-to-gas ratio, local gas density and density gradient.
The stellar feedback is implemented with the standard ``delayed cooling'' model \citep{stinson06}. Briefly speaking, the gas is prevented  to cool for $\sim 10$ Myr near a new star particle.
With these prescriptions, in CROC simulations, halos below $10^8 \Msun$ do not form any stars because they lack molecular gas.

\subsection{Quasar Implementation}
I study the quasar radiative feedback by running a suite of simulations with and without quasars. I start these simulations from the snapshot of CROC simulation B40E at redshift $z=8.9$. In this snapshot, there are two halos with dark matter mass greater than $1\times10^{11}\Msun$. Under the abundance matching ansatz, it is likely that they host quasars of modest magnitude $\M1500 \sim -24$ \citep{chen18}, comparable to the quasar observed by \citet{ota18}.

I choose the most massive halo (dark matter halo mass $M_h=1.6\times10^{11}\Msun$) at this snapshot as the quasar host galaxy. The quasar is modeled as a massless ultraluminous point source. To follow the motion of the quasar, I track 50 old stars in the gas center of the halo. The quasar is then placed at the median position of these stars. By doing so, the quasar remains in the gas center of the galaxy, which is a widely accepted nature of real quasars.

The current RT solver in my simulation requires all the radiation sources to use the same type of spectrum to speed up RT calculations to couple it to hydrodynamic calculations. Therefore the spectrum of the quasar is the same  as stars in the simulations. More specifically, the spectrum has a spectral index of $1.8$ above $1$ Rydberg and an exponential cutoff above $4$ Rydberg \citep[see the solid line in Figure 4 of][]{ricotti02}. The light curve of the quasar is described by a simple light-bulb model, in which they have a fixed luminosity when they are turned on. The quasar's luminosity is controlled by the total luminosity of ionizing photons  $L_Q$. At a given a magnitude $\M1500 $,  $L_Q$ can vary widely depending on the spectra index, which is poorly understood for $z\sim6$ quasars. According to \citet{paris11}, the spectra index for $z<3$ quasars can varies from 1.8 to 0.5. For a $\M1500 =-24$ quasar, this corresponds to $ L_Q = 3\times10^{45}$ to $3\times10^{47} \rm erg/s$. In this study, I use $L_Q=1\times10^{46} \rm erg/s$ and $1\times 10^{47} \rm erg/s$ as the luminosities for low and high luminosity quasars, respectively.

\subsection{Simulation Suite}
I run several simulations to explore the dependency of  quasar radiative feedback on different variables , as shown in Table \ref{Table:sim}. NoQ simulation is the reference run without quasar. Q\_z9\_L runs with a low luminosity quasar turned on for $309$ Myr from $z=8.9$ to $z=6.4$, a relatively long quasar lifetime. Q\_z9\_H is the same as Q\_z9\_L except that the quasar is 10 times brighter. This simulation is expected to show the maximum effect of quasar radiative feedback on galaxy formation.
In simulation Q\_z7\_H I do not turn on the bright quasar until $z=6.8$. By comparing this simulation with Q\_z9\_L and Q\_z9\_H, we can study the relative importance of quasar lifetime and quasar luminosity on radiative feedback. In the last simulations Q\_z9\_L\_z8off and Q\_z9\_L\_z7off I turn off the quasar at $z=8.2$ and $z=6.8$, respectively. These simulations allow us to study the behavior of galaxy formation in the ``post-quasar'' phase.

In Figure \ref{fig:xHI} I show the neutral hydrogen map of three simulations NoQ, Q\_z9\_L and Q\_z9\_H at $z=8.6$, centered on the quasar host halo. In the left panel where the quasar is not turned on, ionized bubbles are relatively small and clustered around the massive halo. While in the middle and right panels, the quasar radiation dominates near the quasar host. At this time the quasar has been shining for $t_Q=30$ Myr, and the ionized bubble in the  Q\_z9\_H run (right panel) has grown to a radius $\gtrsim 1 $ pMpc.

\begin{table}[t]
\centering
\caption{Simulation Parameters }
\begin{tabular}{lrrc}
\hline
\hline
Sim Name        & $L_Q$ [erg/s] & Turn-on Epoch & legend color\\
NoQ             & -          & -        & blue     \\
Q\_z9\_L        & $1\times10^{46}$    &  $z=8.9 - z=6.4$ & orange\\
Q\_z9\_H        & $1\times10^{47}$       & $z=8.9 - z=6.4$ &red\\
Q\_z7\_H        & $1\times10^{47}$       & $z=6.8 - z=6.4$ &purple\\
Q\_z9\_L\_z8off & $1\times10^{46}$       & $z=8.9 - z=8.2$ & black\\
Q\_z9\_L\_z7off & $1\times10^{46}$       & $z=8.9 - z=6.8$ &brown\\

\hline 

\label{Table:sim}
\end{tabular}
\end{table}

\subsection{Halo Catalog and Halo Properties}
I use HOP \citep{eisenstein98}, a built-in halo finder in yt\footnote{http://yt-project.org/} \citep{turk11}, to generate halo catalogs. I use the default density threshold of 160 to obtain dark matter halo mass $M_{h}$ and halo radius $r_h$. At above $5\times10^8 \Msun$, the halo catalog is complete.

To quantify the quasar radiation feedback to a specific halo, it is useful to identify the ``same'' halos across simulations. This is done by matching the positions and mass of halos in two catalogs. If the distance between two halos (in two halo catalogs respectively) are less than 50 comoving kpc and their halo mass difference is less than 10\%, they are identified as the same halo.

I then calculate halo properties like gas mass and gas temperature within $r_h$ for each halo, as well as the 
 unattenuated galaxy magnitude $\M1500 $. $\M1500 $ is calculated using the Starburst99 model \citep{leitherer99}. Specifically, with information of age, metallicity and mass of a star particle, I search a Starburst99 look-up table for its luminosity at $1500\mathring{\rm A}$, and sum the luminosity for all star particles within $r_h$ of each halo. Finally, I convert the luminosity to AB magnitude. With the information of mass and age of star particles in each halo in the last snapshot, I also obtain the full star formation history for each halo.

\begin{figure}
\centering
\includegraphics[width=0.45\textwidth]{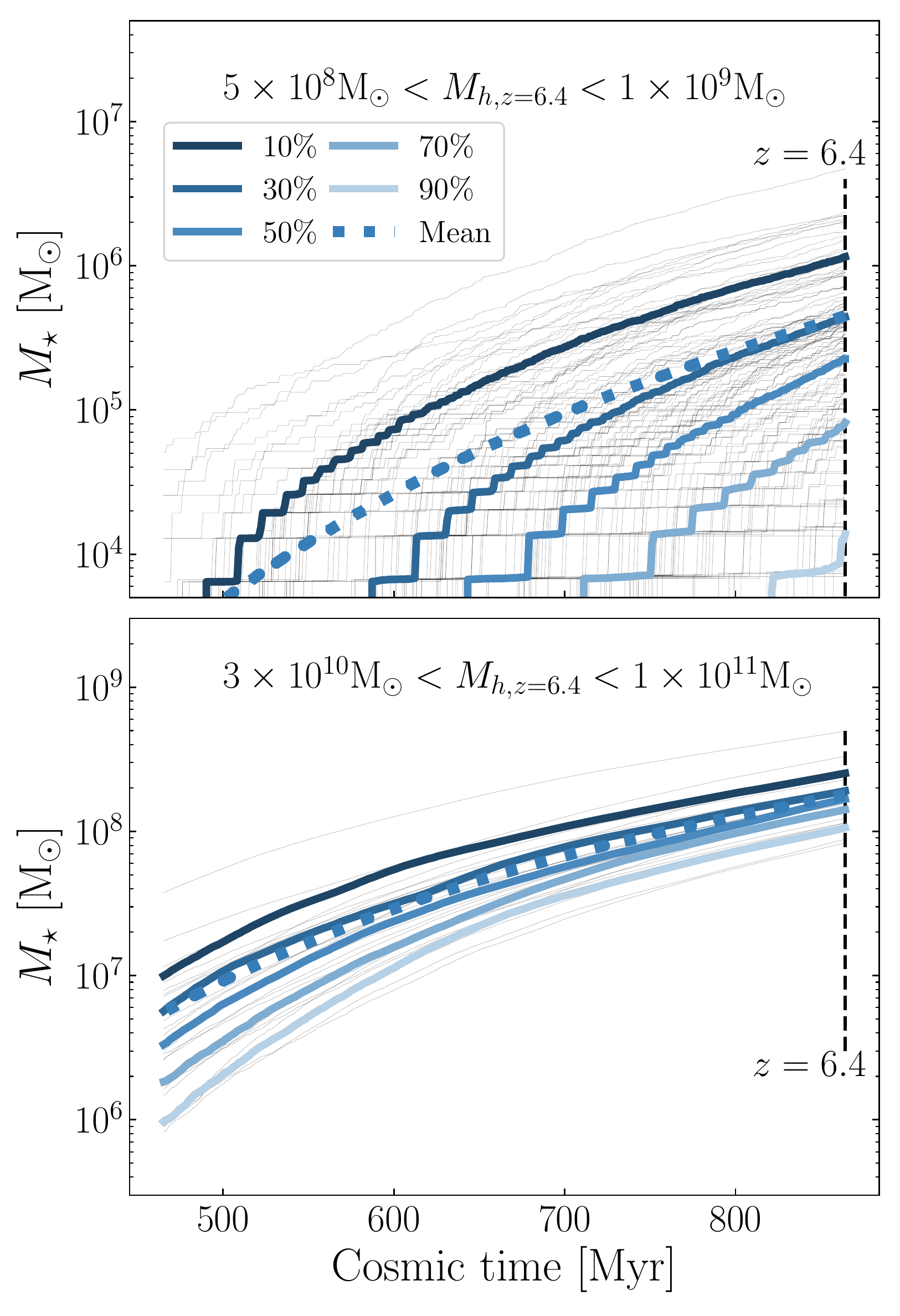}
\caption{Star formation histories (SFHs) of low mass halos (upper panel) and high mass halos (lower panel) in no quasar simulation NoQ. 
Halos are selected within a $1$ pMpc sphere centered on the quasar host at z=6.4. Black thin lines are SFHs in individual halos. In the upper panel, the number of individual halos are down-sampled by a factor of 10. The lower panel shows SFHs for all individual halos.
The thick blue lines show a certain percentile of stellar mass at each time. The dashed blue lines are the stellar mass averaged over all the halos in the halo mass bins, which is used to calculate the star formation rates (SFRs).
}\label{fig: SFHvariety}
\end{figure}

\begin{figure}
\centering
\includegraphics[width=0.45\textwidth]{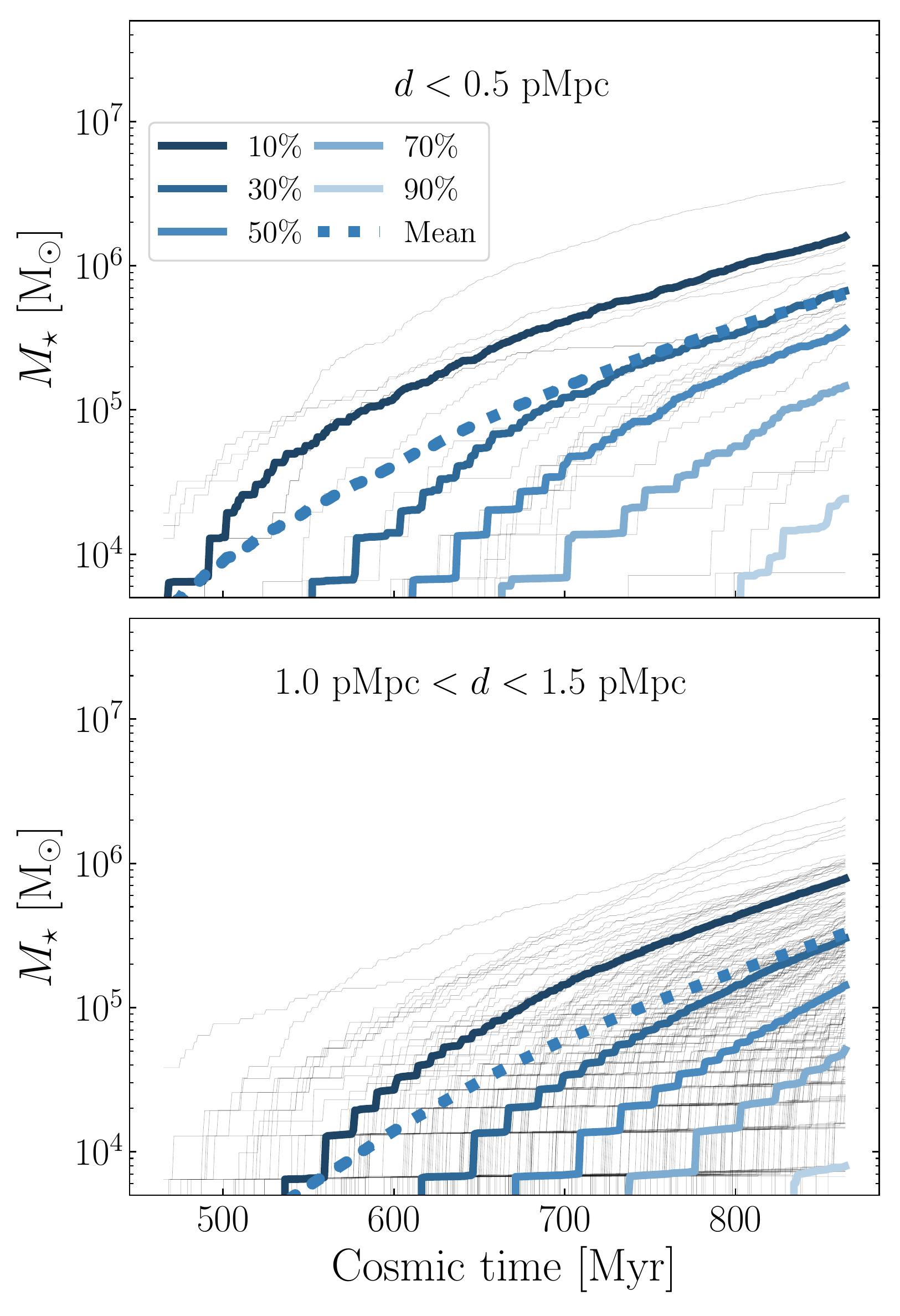}
\caption{SFHs of low mass halos with $5\times10^{8}\Msun-1\times10^{9}\Msun$ similar to the upper panel of Figure \ref{fig: SFHvariety}, but grouped into different distance bins to the quasar host.
Upper panel: halos within $ 0.5 \rm pMpc$ from the quasar host. Lower panel: halos within $1.0 {\rm pMpc} < d < 1.5 \rm pMpc$ from the quasar host. 
There is a distance dependence in SFH for these low mass halos.}\label{fig:SFHdistance}
\end{figure}

\section{Star Formation History and Gas Properties}\label{sec:galaxyProp}

\subsection{SFHs in dense environment}

\begin{figure*}
\centering
\includegraphics[width=0.8\textwidth]{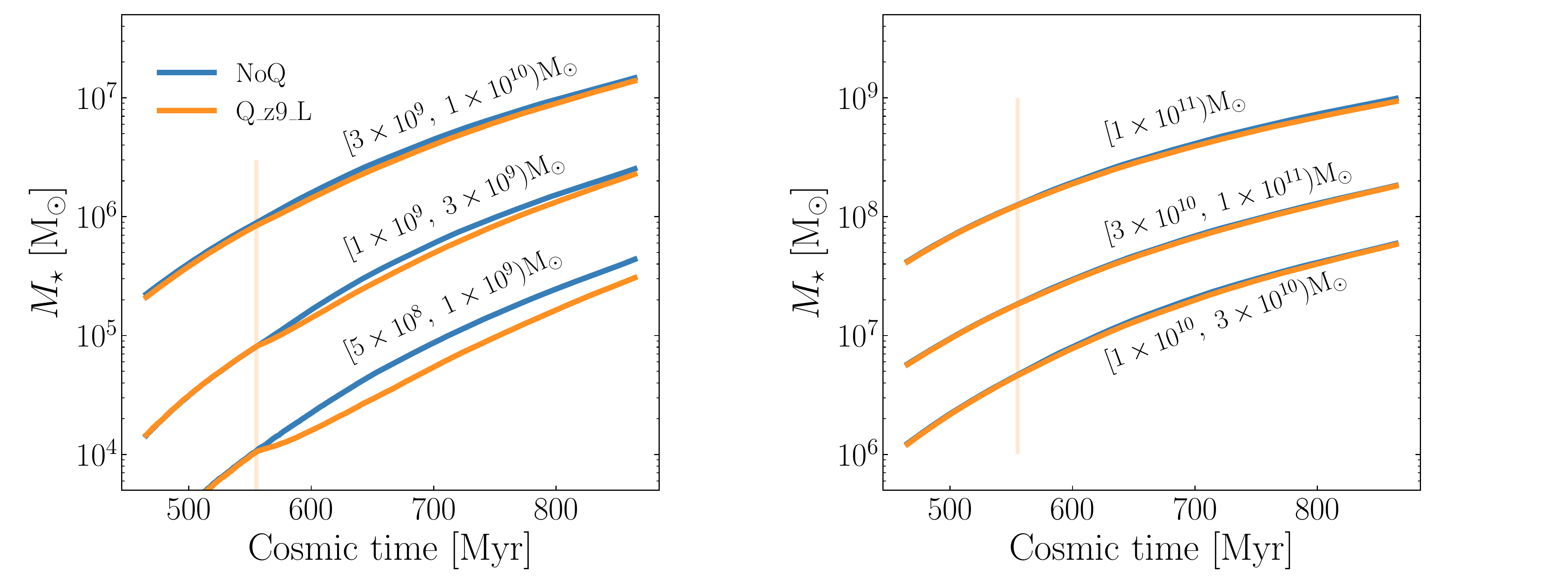}
\caption{Evolution of the mean stellar mass in six halo mass  bins. Blue lines are for halos in NoQ run and orange for Q\_z9\_L run. All these halos are from a sphere of $1$ pMpc centered on the quasar host. The light orange lines mark the time when the quasar turns on in Q\_z9\_L run.}\label{fig: QSO_SFH}
\end{figure*}

\begin{figure*}
\centering
\includegraphics[width=0.8\textwidth]{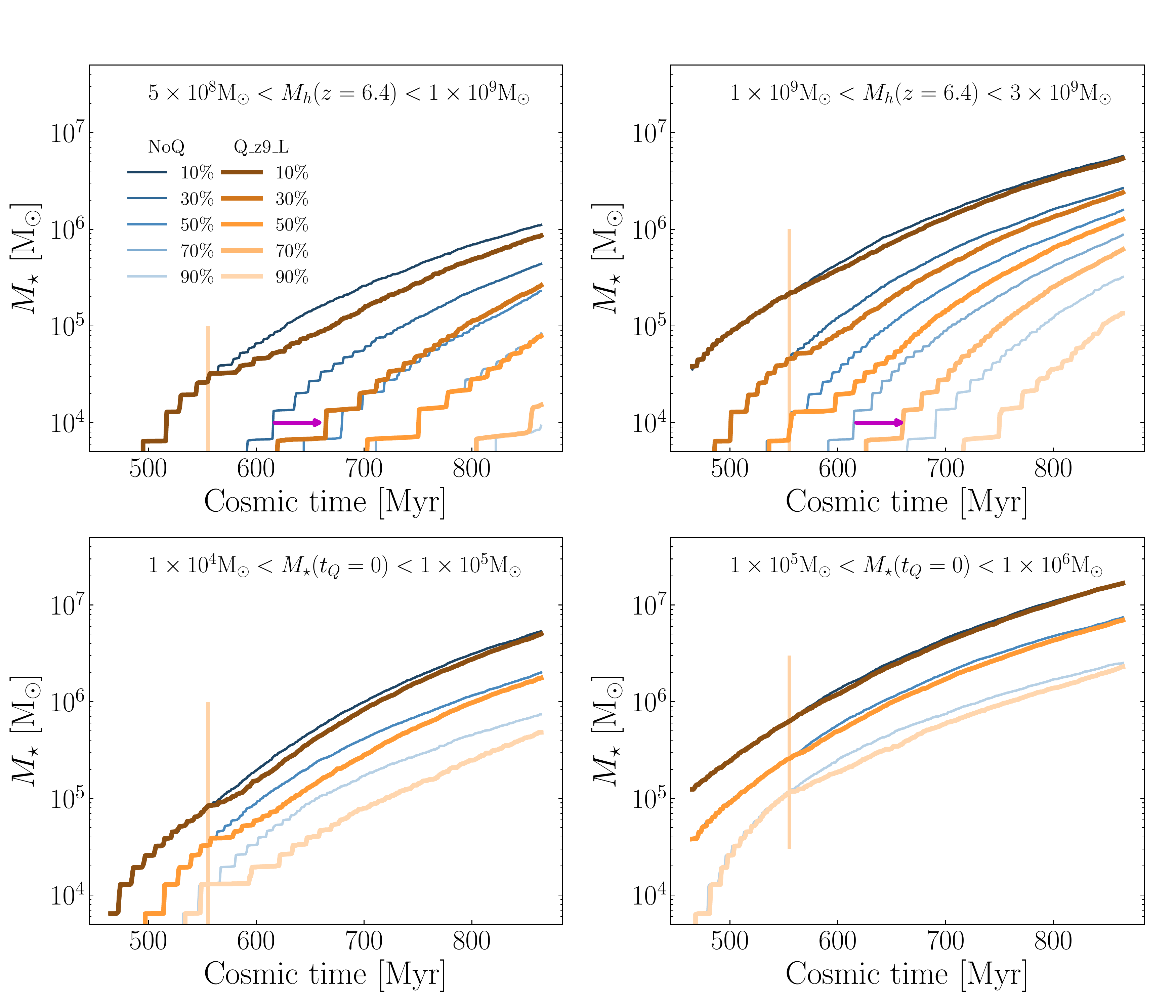}
\caption{Comparison of percentile SFHs between no quasar simulation NoQ and simulation with the low luminosity quasar Q\_z9\_L. 
Upper panels: halos are grouped by their halo mass at z=6.4, within a $1$ pMpc sphere of the quasar host.
Upper left: Halos with mass $5\times10^{8}\Msun-1\times10^{9}\Msun$. 
Upper right:  Halos with mass $1\times10^{9}\Msun-3\times10^{9}\Msun$.
The thin blue lines are the same as Figure \ref{fig: SFHvariety}, showing the percentile SFHs in NoQ simulation. 
The thick orange lines show SFHs in quasar simulation Q\_z9\_L. 
The vertical orange line marks the time when the quasar is turned on in simulation Q\_z9\_L. 
After the quasar turns on, every percentile SFHs lines in Q\_z9\_L shifts to the right of the corresponding blue lines, as indicated by the magenta arrows.
Note that the 90\% line of the quasar simulation disappears in the left panel because these halos have not formed any stars. 
Lower panels: Similar to the upper panels, but halos are grouped by their stellar mass at z=8.9, the moment the quasar just turns on in Q\_z9\_L run. 
Lower left: Halos with stellar mass $1\times10^{4}\Msun-1\times10^{5}\Msun$ at z=8.9.
Lower right: Halos with stellar mass $1\times10^{5}\Msun-1\times10^{6}\Msun$ at z=8.9.
}
\label{fig:QSO_compare}
\end{figure*}
If we look at how diverse star-formation histories are in the simulation without quasars, we will later better appreciate how quasars impact the behaviors of star formation.
In this section, I use no quasar simulation NoQ to study this ``intrinsic'' SFHs. 
Specifically, I examine the SFHs of halos grouped at z=6.4 into six halo mass bins ( $5\times10^{8}\Msun-1\times10^{9}\Msun, 1\times10^{9}\Msun-3\times10^{9}\Msun, 3\times10^{9}\Msun-1\times10^{10}\Msun, 1\times10^{10}\Msun-3\times10^{10}\Msun,3\times10^{10}\Msun-1\times10^{11}\Msun,> 1\times10^{11}\Msun$).


Halos have very different SFHs, even for those in a narrow halo mass bins. This is especially the case for low mass halos, as shown in the upper panel of Figure \ref{fig: SFHvariety}. The halos shown here are from a sphere of $1$ pMpc centered on the quasar host at $z=6.4$.
In the upper panel, I plot the SFHs of halos in the lowest halo mass bin.  
Black thin lines are SFHs for individual halos. I randomly select a tenth of the halos to plot, due to the large number of small mass halos. The blue lines mark certain percentiles of the stellar mass at each time. 
The scatter (10\% to 90\%) in stellar mass is as large as 2 dexes in these low mass halos. 
As halo mass increases, this scatter decreases. In the high halo mass bin $3\times10^{10}\Msun-1\times10^{11}\Msun$, the scatter reduced to ($\sim 0.5$ dex), as shown in the lower panel of Figure \ref{fig: SFHvariety}. 
The larger scatter of stellar mass in lower mass halos could be due to environmental effects. For example, there are more subhalos in lower halo mass bins which accrete mass early but halt later, thus have a higher concentration and more stars formed in the center.


I also calculate the mean stellar mass averaging over all halos in each halo mass bin (dashed line). We can see the stellar mass - halo mass ratio $M_\star/M_h$ at $z=6.4$ increases from 0.0007 for the lowest halo mass bin $5\times10^{8}\Msun-1\times10^{9}\Msun$ to 0.004 for higher halo mass bin $3\times 10^{10} - 1\times10^{11}$. This is consistent with \citep{behroozi18} and is a natural consequence that higher mass halo has higher gas fraction \citep{gnedin14b}.
I briefly note that because of the order of magnitude difference in stellar mass, the mean stellar mass does not trace the median of halo mass, especially in the low halo mass bins (upper three panels of Figure \ref{fig: SFHvariety}). 
Halos with larger stellar mass drive the mean stellar mass, and because they have smoother evolution, the mean stellar mass evolution also shows a smoother behavior than the percentile stellar mass evolution.
When study EoR, we are usually more interested in the total stars formed. Therefore, the mean stellar mass is more appropriate to use when calculate the SFR in each halo mass bins. In the following sections SFHs are calculated by differentiating the mean stellar mass evolution.


We have seen that the low mass halos have a large variation in SFHs, which may be a result of environmental effects. This environmental dependence should also lead to a systematic difference in SFHs between halos close to the quasar and far away from the quasar. In Figure \ref{fig:SFHdistance}, I plot the SFHs of low mass halos with $5\times10^{8}\Msun-1\times10^{9}\Msun$ into different distance bins. The upper panel shows the halos within $0.5$ pMpc, while the lower panel shows the halos in a shell $1.0 - 1.5$ pMpc from the quasar host. Closer to the quasar host, halos tend to form stars earlier and eventually have more stars. This is partially because compare to normal places, in the denser environment subhalo fraction is higher. These subhalos have early accretion which leads to a higher concentration, thus a denser gas core where stars are efficiently formed. The SFHs for halos beyond 1.5 pMpc is very similar to the distance bin 1 - 1.5 pMpc. This distance dependence is weaker for more massive halos. In halo mass bins larger than $3\times10^{9}\Msun$ the distance dependency diminishes to nearly unnoticeable.

\subsection{The Impact of Quasar Radiations on SFHs} \label{sec:QL}
Quasar radiation can impact star formation in surrounding galaxies, but the level of impact depends on many factors, the most prominent two of which are halo mass and radiation intensity. In this subsection I focus on how quasar radiative feedback impacts SFHs in halos with different halo mass, using simulations NoQ and Q\_z9\_L. In the next section, I will study the dependence on quasar luminosity.

\begin{figure*}[!htp]
\centering
\includegraphics[width=\textwidth]{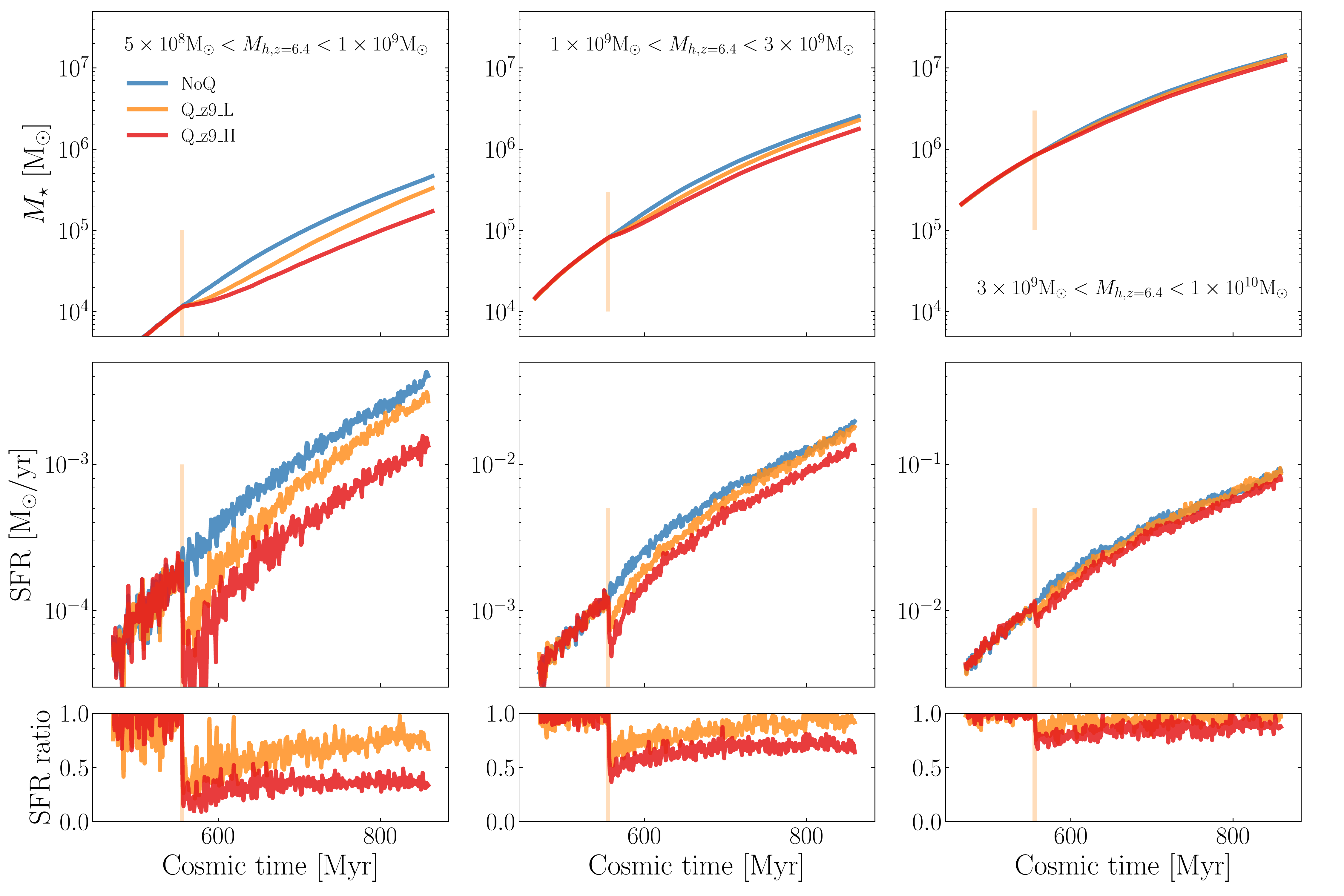}
\caption{Upper panels: mean SFHs of halos in three low mass bins. Halos are within a sphere of $1$ pMpc from the quasar host. Blue, orange and red lines represent SFHs in simulations NoQ, Q\_z9\_L and Q\_z9\_H, respectively. Middle panels: SFRs as a function of time in three low halo mass bins. Lower panels: star formation rate ratio compared with the simulation NoQ. The vertical light orange lines mark the time when quasar turns on.}\label{fig: QSO_SFR}
\end{figure*}

\begin{figure*}[!htp]
\centering
\includegraphics[width=\textwidth]{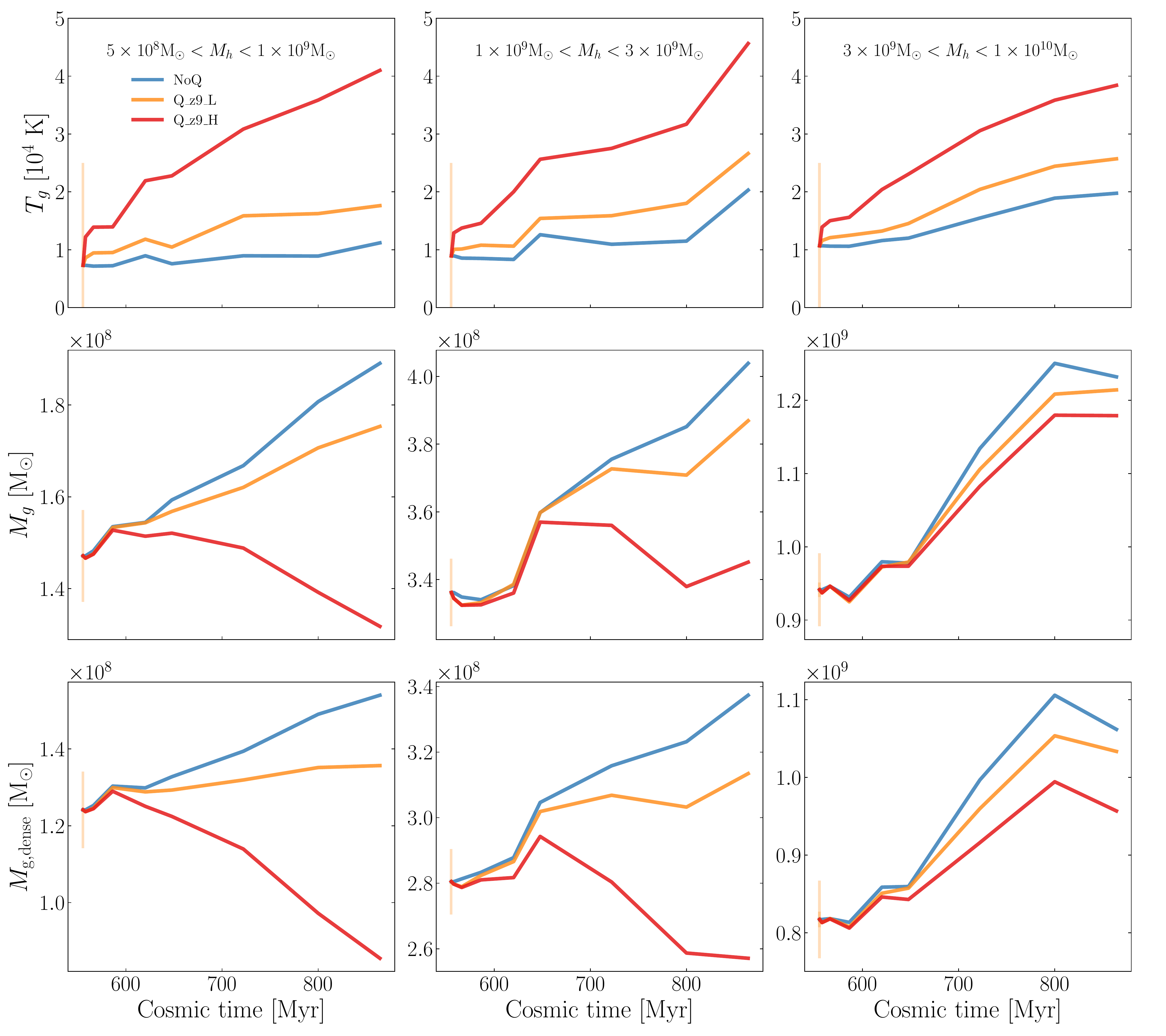}
\caption{Gas properties evolution for simulation NoQ (blue), Q\_z9\_L (orange) and Q\_z9\_H (red). From top to bottom: gas temperature, total gas mass and dense gas mass. Dense gas is defined as density above $3\times10^{-24} ~\rm~g~cm^{-3}$. Different columns are halos in different mass bins. All these halos are from a sphere of $1$ pMpc centered on the quasar host.}\label{fig: gas_prop}
\end{figure*} 

In Figure \ref{fig: QSO_SFH}, I group halos within 1 pMpc of the quasar host into six mass bins and plot the mean SFHs in simulations with and without the quasar. Halos in the lowest halo mass bin  $5\times10^{8}\Msun-1\times10^{9}\Msun$ show significant suppression of star formation. As the halo mass increases, the differences between different simulations decrease. For halos larger than $3\times10^{9}$, there is no significant suppression. 

It is possible that the quasar only suppresses star formation in halos with very low stellar mass (lighter blue lines in Figure \ref{fig: SFHvariety}) that causes the suppression we see in Figure \ref{fig: QSO_SFH}. To examine this, I further plot the percentile SFHs of simulations NoQ and Q\_z9\_L, which is shown in the upper panels of Figure \ref{fig:QSO_compare}. For the lowest mass bin $5\times10^{8}\Msun-1\times10^{9}\Msun$ (upper left panel), we can see a systematic delay of SFHs for halos which form stars only after the quasar turns on, as pointed out by the magenta arrow. Note that the 90\% line of the quasar simulation disappears in the left panel because these halos have not formed any stars. In halos that have already formed stars before quasar turns on (e.g., the 10\% line), quasar suddenly suppresses the ongoing star formation. For halos that form stars only after the quasar turn-on time, the quasar delays the star formation. 

Interestingly, we can find some similarities across different halo mass bins. For example, the delay of the $30\%$ line in the upper left panel and the $70\%$ line in the upper right panel are similar, and so are the $10\%$ line in the upper left panel and the $30\%$ line in the upper right panel. This indicates the stellar mass at $t_Q=0$ is a better proxy to the degree of suppression.
In the lower panels of Figure \ref{fig:QSO_compare}, I regroup the halos according to their stellar mass at $t_Q=0$. In the lower left panel I show halos with stellar mass between $1\times10^{4}\Msun-1\times10^{5}\Msun$ and the lower right $1\times10^{5}\Msun-1\times10^{6}\Msun$. I omitted 30\% and 70\% lines for clarity.
We can see that for halos with $M_\star<10^5\Msun$, there is a visible delay in star formation. However, for higher stellar mass halo this delay diminishes and totally disappears when $M_\star>10^6\Msun$.

\begin{figure}
\centering
\includegraphics[width=0.5\textwidth]{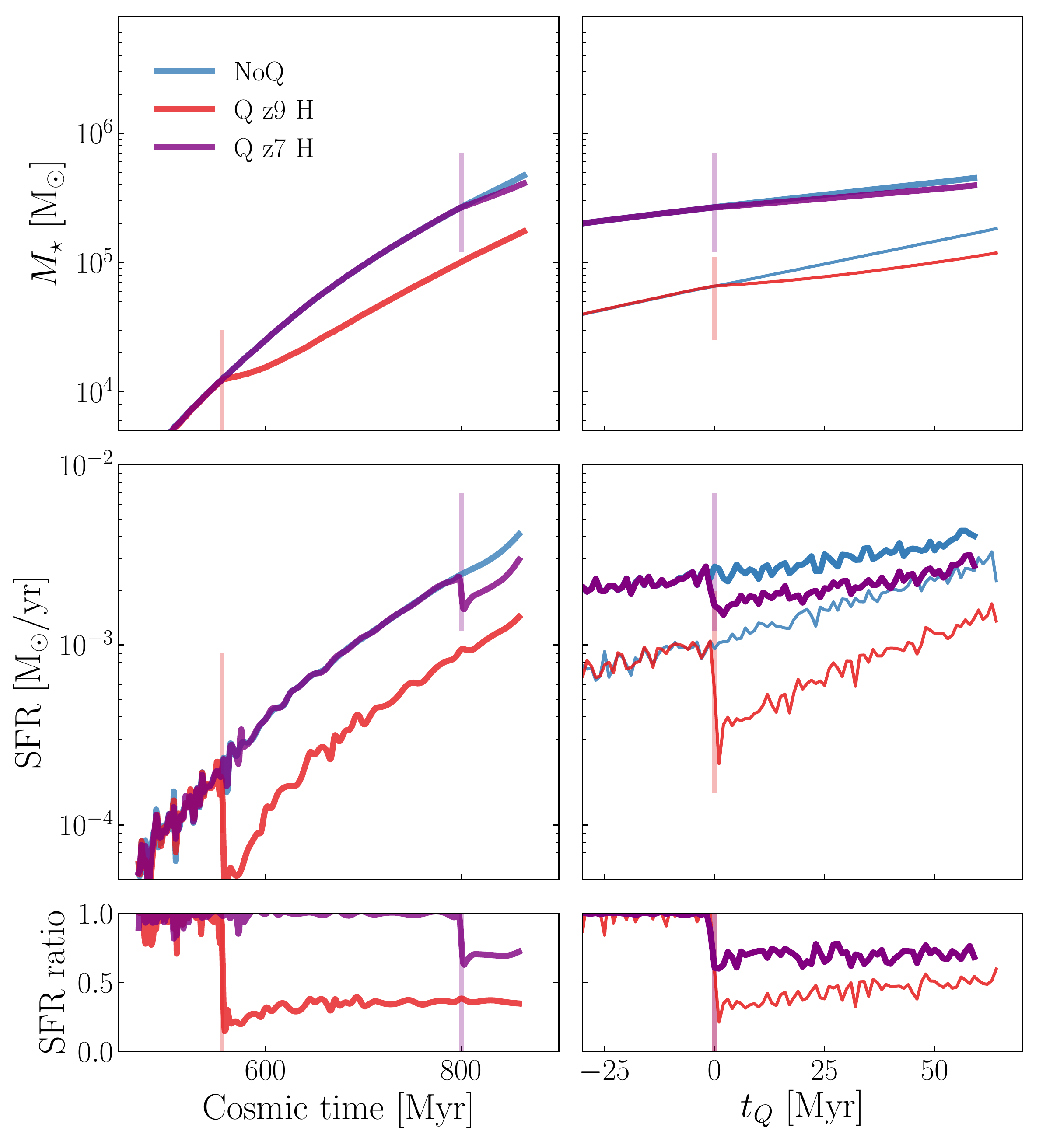}
\caption{Left: SFHs (top), SFRs (middle) and SFR ratios (bottom) same as the left column of Figure \ref{fig: QSO_SFR}, but for simulations NoQ (blue), Q\_z9\_H (red) and Q\_z7\_H (purple), where the quasar turns on later, marked by the vertical purple line. Right: Similar to the left column, but halos are grouped by their halo mass at the same quasar lifetime $t_Q=64$ Myr, instead of the same cosmic time.}\label{fig: QSO_late_turnon}
\end{figure}

\begin{figure}
\centering
\includegraphics[width=0.45\textwidth]{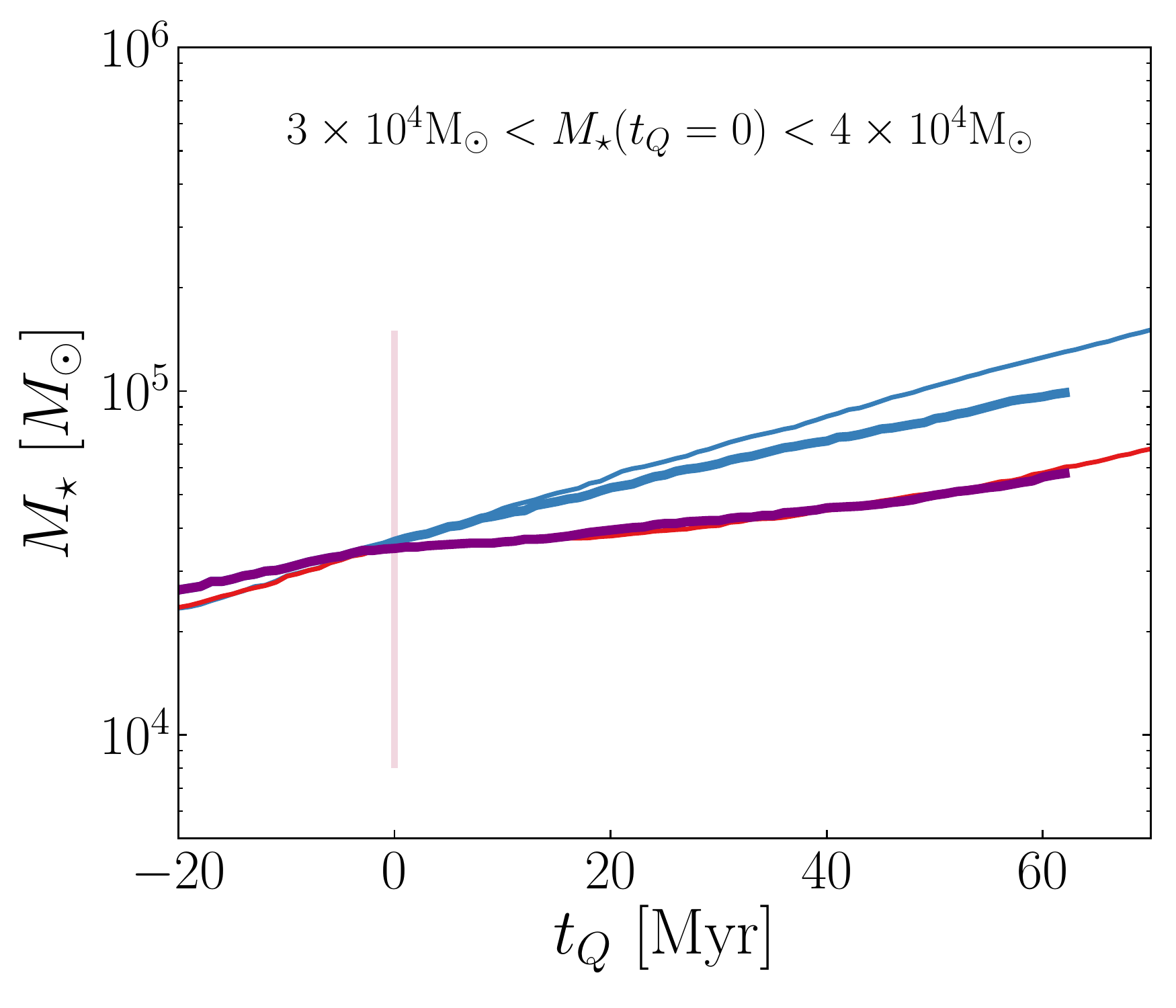}
\caption{The mean SFHs of halos in simulations Q\_z9\_H (thin red) and Q\_z7\_H (thick purple). The x-axis here is the quasar lifetime. The halos are selected within $1$ pMpc around the quasar host and have stellar mass between $3\times10^4 \Msun$ and $4\times10^4\Msun$ at $t_Q=0$. For comparison, the two blue lines are halos in simulation NoQ: the thin blue line represents the same halos as the ones in  Q\_z9\_H (thin red line); the thick blue line represents the same halos as the ones in  Q\_z7\_H (thick purple line). 
}\label{fig: QSO_late_turnon_groupby_Ms}
\end{figure}


\subsection{Dependence on quasar luminosity}
We expect quasar luminosity impacts the degree of change in star formation of halos in quasar fields. 
In this section, I analyze three simulations NoQ, Q\_z9\_L, and Q\_z9\_H to study this dependency on quasar luminosity. In the upper panels of Figure \ref{fig: QSO_SFR}, I plot the mean SFHs of  halos in three halo mass bins $5\times10^{8}\Msun-1\times10^{9}\Msun$ (left), $1\times10^{9}\Msun-3\times10^{9}\Msun$ (middle) and $3\times10^{9}\Msun-1\times10^{10}\Msun$ (right) in these three simulations.
The red line shows the SFHs of the high luminosity run Q\_z9\_H where the quasar is 10 times brighter than that in Q\_z9\_L. It shows that higher luminosity quasar has a greater impact on SFHs than the low luminosity quasar, especially in low mass halos. 

In observation, SFRs are more accessible, so in the middle panels, I plot the star formation rates. 
We notice that the suppression happens right after the quasar turns on. This short timescale indicates the main suppression mechanism is photo-dissociation rather than photo-evaporation, because the latter operates in the timescale of $\sim$ 100 Myr \citep{shapiro04}.
In our simulation, star formation rate is directly proportional to the mass of molecular hydrogen, which depends on the LW radiation. When the quasar turns on, the LW radiation immediately increases, causing the suppression of star formation rate.

In the lower panels, I show the ratio of SFRs of the two quasar runs to NoQ. Within $100$ Myr after the quasars turn on, the SFR ratio only differs by less than a factor of two, although the luminosity of quasar has an order of magnitude difference. This weak dependence is largely due to the complex shielding of dust and the self-shielding of molecular hydrogen, which in our simulation is done using the subgrid model of \citet{gnedin14c}. In Section \ref{sec:discussion} I will discuss it in more detail.

Although photo-heating does not impact star formation rate immediately, it alters the property of gas in halos and has a long-term effect on star formation history, as we will better appreciate later in Section \ref{sec:turnoff}, when I show the quasar turn-off simulations.  In this section, I show how the quasar radiation alters the gas properties of low mass halos in simulation NoQ, Q\_z9\_L, and Q\_z9\_H. In the upper panel of Figure \ref{fig: gas_prop} I select halos within 1 pMpc from the quasar host and show the averaged gas temperature evolution for three low halo mass bins. When close to the quasar, the ionization front moves very fast and heats the gas very efficiently. Therefore, soon after the quasar turns on, it raises gas temperature significantly. Also, Q\_z9\_H run shows a higher gas temperature than Q\_z9\_L. This may be surprising because \citet{abel99} has shown that the gas heating does not depend on radiation intensity explicitly, but rather on  radiation spectrum -- harder spectrum heats the gas more efficiently than softer spectrum. The dependency on quasar luminosity we see here is mostly due to the higher luminosity quasar creating faster ionization fronts, which travel deeper inside the halos and heat more gas. Because of the higher optical depth inside the halo, the spectrum also becomes harder and heats the gas more efficiently \citet{abel99}. In the meantime, the recombination rate decreases as temperature increases, which further reinforces heating.

In the middle panels, I show the evolution of total gas mass. As a consequence of the increased gas temperature, halo mass in the quasar simulations are significantly lower than NoQ run after $\sim 100$ Myr, roughly the sound-crossing time of these halos. This decrease in total gas mass depends on the halo mass. The smaller the halo, the faster and more significant the decrease is. Especially for the lowest mass bin in Q\_z9\_H, the halos suffer from net gas loss, or ``photo-evaporation''. In contrast, for halos more massive than $3\times10^9 \Msun$, the decrease is not significant.

Similar to the middle panel, in the lower panels of Figure \ref{fig: gas_prop} I plot the mass of dense gas, defined as $n_g>3\times10^{-24} \rm ~g~cm^{-3}$, which is slightly above the density threshold (0.01 $\rm cm^{-3}$) that stars can form in our simulation. For the lowest mass bin, the average dense gas mass in Q\_z9\_H is only 75\% of that in NoQ after $\sim$ 200 Myr. This shows that photo-heating is also an important mechanism to suppress star formation at long time scales.

\subsection{Case of late time turn-on}

If a quasar is on for a shorter period of time, we expect it to impact star formation less. In this section I compare simulations NoQ, Q\_z9\_H and Q\_z7\_H to study the star formation dependency on different quasar lifetimes.

In the left column of Figure \ref{fig: QSO_late_turnon}, I plot the SFHs of low mass halos ($5\times10^{8}\Msun-1\times10^{9}\Msun$) in these three simulations in the top panel. Similar to Figure \ref{fig: QSO_SFR}, I group the halos within 1 pMpc from the quasar according to their halo mass at $z=6.4$, or cosmic time = 864 Myr. At this time the quasar in Q\_z9\_H has been on for 309 Myr and that in Q\_z7\_H for 64 Myr. It is clear that in Q\_z7\_H, the total suppression at $z=6.4$ is significantly less than that in Q\_z9\_H. Not only the total suppression, but also the instantaneous suppression when the quasar just turns on, is significantly less, as manifested by the milder drop in SFRs shown in the middle panels. The SFR ratios of the two quasar runs relative to the NoQ run are shown in the bottom left panel. For these low mass halos, the SFR ratio is only $\sim$0.3 for the early turn on simulation, compared to 0.7 for the late turn on simulation.

This large difference in the suppression of SFR is partially because by the time the quasar turns on at $z=6.8$, halos have grown significantly. Most of the halos have doubled their halo mass during the 309 Myr from $z=8.9$ to $z=6.8$. To eliminate the halo growth and study other impact factors, I regroup the halos by their halo mass at the same quasar lifetime, rather than the same cosmic time. In the right column of Figure \ref{fig: QSO_late_turnon}, I plot the same quantities as the left column, but for halos with $5\times10^{8}\Msun-1\times10^{9}\Msun$ at the same quasar lifetime $t_Q=64$ Myr for both simulation Q\_z9\_H (thin red) and Q\_z7\_H (thick purple). The thin (thick) blue lines, as references, are from simulation NoQ at the same time as the thin red (thick purple) lines. Interestingly, comparing the thin and thick lines of this figure, we can find even for halos with the same halo mass, at lower $z=6.4$, the average stellar mass is twice as much as those at $z=8.7$. In Section \ref{sec:QL}, we have seen that quasar radiation has a very limited impact on high stellar mass halos. The result here is similar: Since the stellar mass of $z=6.4$ halos are larger, the suppression from quasar radiation is smaller, as shown in the middle and lower panels.

In Figure \ref{fig: QSO_late_turnon_groupby_Ms}, I examine what the suppression looks like if I group halos by their stellar mass at $t_Q=0$. Specifically, in a $1$ pMpc sphere centered on the quasar host in simulation NoQ, I select all the halos with stellar mass between $3\times10^4 \Msun$ and $4\times10^4 \Msun$ at z=8.9 (marked by the vertical purple line). Their mean SFH is shown as the thin blue line. Then I select the same halos in the Q\_z9\_H run, and their mean SFH is shown as the thin red line. 
Similarly, at a later time z=6.8 (also marked by the vertical red line), I select halos with the same stellar mass range ($3\times10^4 \Msun - 4\times10^4 \Msun$) in simulation NoQ. Their mean SFH is shown as the thick blue line. Then I select the same halos in the Q\_z7\_H run, and their SFHs are shown as the thick purple line. We can find that the thick purple line overlaps very well with the thin red line. This demonstrates that stellar mass at $t_Q=0$ is a good indicator of the degree of suppression: Even the quasar turns on at two different redshifts $8.9$ and $6.4$ separated by $245$ Myr, at a fixed quasar lifetime, halos with the same stellar mass experience the similar degree of suppression.

\subsection{Recovery after the quasar turns off}
\label{sec:turnoff}

It is possible that quasars change the halo properties so significantly that even after the quasar turns off, they still leave a significant impact. In this section, by comparing simulations NoQ, Q\_z9\_L, Q\_z9\_L\_z8off and Q\_z9\_L\_z7off, I will show that if the quasar turns off early ($t_Q \sim 60$ Myr), the star formation will recover quickly, while if the quasar shines over 200 Myr the star formation will recover slowly.

\begin{figure}
\centering
\includegraphics[width=0.238\textwidth]{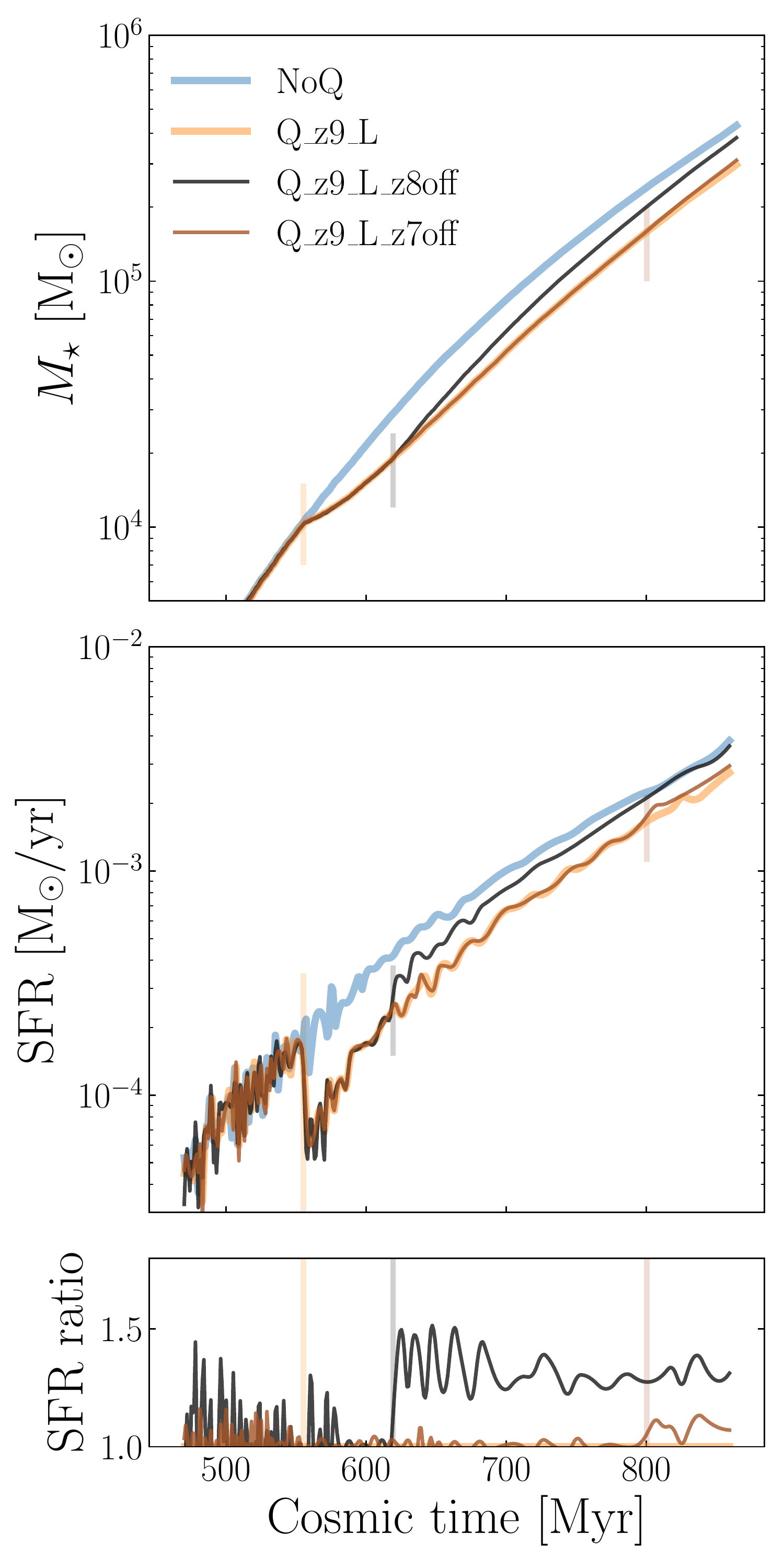}
\includegraphics[width=0.235\textwidth]{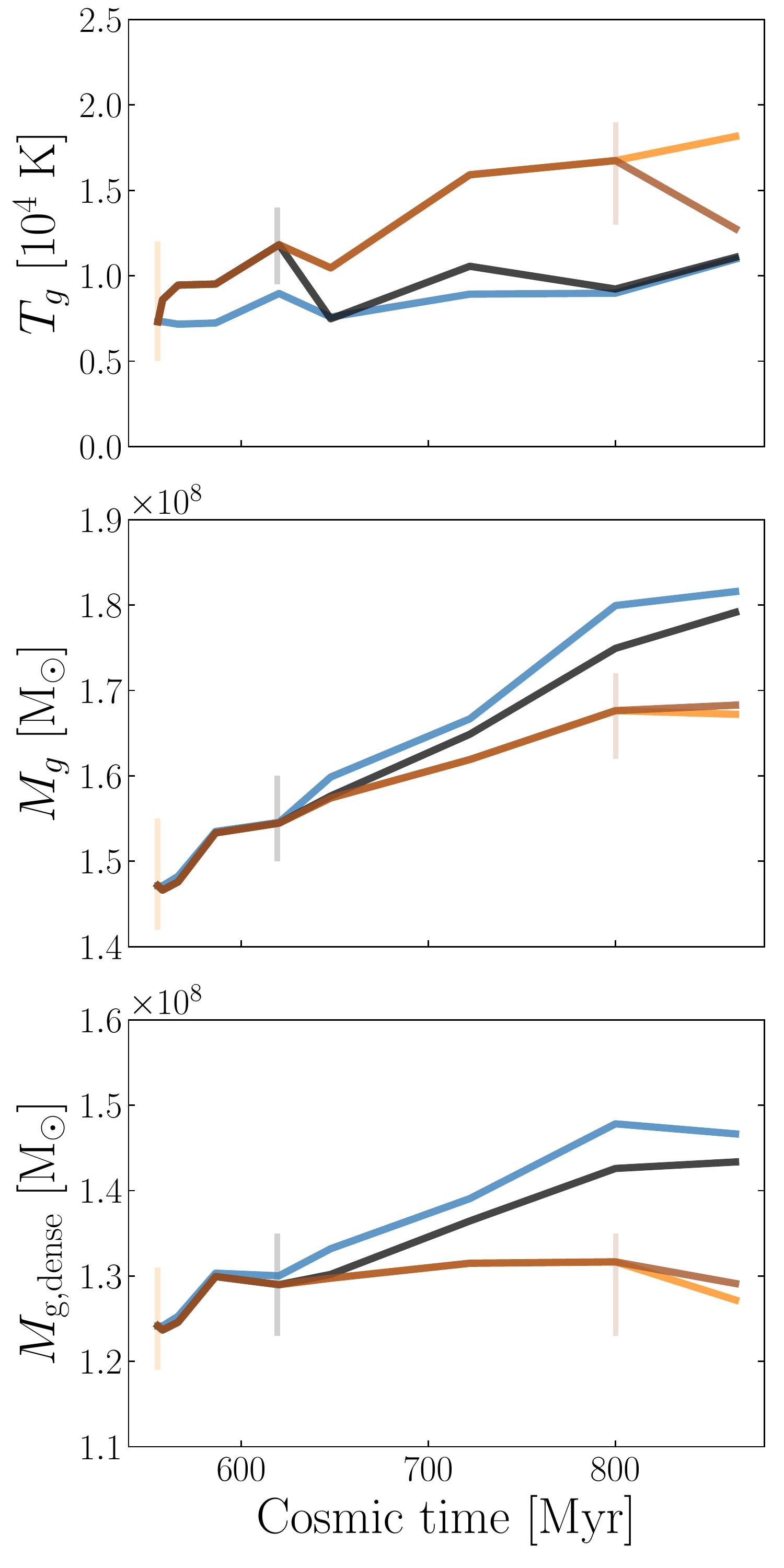}
\caption{Left: SFHs (top) and SFRs (middle), similar to the left column of Figure \ref{fig: QSO_SFR}, but for simulations NoQ (blue), Q\_z9\_L (orange),  Q\_z9\_L\_z8off (black) and Q\_z9\_L\_z7off (brown). The vertical orange line is when the quasar turns on in the three quasar simulations. The vertical grey line is when the quasar turns off in simulation Q\_z9\_L\_z8off, and the vertical brown line is when the quasar turns off in simulation Q\_z9\_L\_z7off. SFR ratios (bottom) simulations are Q\_z9\_L\_z8off and Q\_z9\_L\_z7off relative to the Q\_z9\_L run. Right: Gas temperature (top), gas mass (middle), and dense gas mass (bottom) similar to the left column of Figure \ref{fig: gas_prop}, but for simulations NoQ (blue), Q\_z9\_L (orange),  Q\_z9\_L\_z8off (black) and Q\_z9\_L\_z7off (brown).}\label{fig: turnoff}
\end{figure}

Similar to the left columns of Figure \ref{fig: QSO_late_turnon}, I analyze the low mass halos ($5\times10^{8}\Msun-1\times10^{9}\Msun$) within $1$ pMpc of the quasar host in simulations NoQ, Q\_z9\_L, Q\_z9\_L\_z8off, and Q\_z9\_L\_z7off.
In the left column of Figure \ref{fig: turnoff}, I show their SFHs. In the middle panels, I plot the SFRs and in the lower panels, I plot the ratio of SFR to the reference Q\_z9\_L run. When the quasar turns off after 64 Myr, the SFR quickly rises, and after $\sim 200$ Myr, it finally catches up to the NoQ simulation where there is no early quasar phase. However, in simulation Q\_z9\_L\_z7off where the quasar has shone for more than 200 Myr, the SFR only slightly increases, as shown by the brown line in the lower panel. 

This dependency of recovery speed on the quasar lifetime is most likely due to the degree of photo-heating. In the right column of Figure \ref{fig: turnoff}, I show the gas temperature, total gas, and dense gas of low mass halos in these four simulations. When the quasar turns off at $z=8.2$ (black lines), the gas temperature soon decreases to be comparable to the NoQ simulation (blue lines). As a consequence, the total gas mass and the dense gas mass do not differ much from the NoQ simulation. However, if the quasar turns off after $\sim$ 200 Myr (brown lines), although the temperature drops significantly within $\sim 60$ Myr, the total gas and dense gas are still almost indistinguishable from the Q\_z9\_L case where the quasar keeps shining. This explains the slow recovery of star formation in the Q\_z9\_L\_z7off simulation.

\begin{figure*}
\centering
\includegraphics[width=5.6cm]{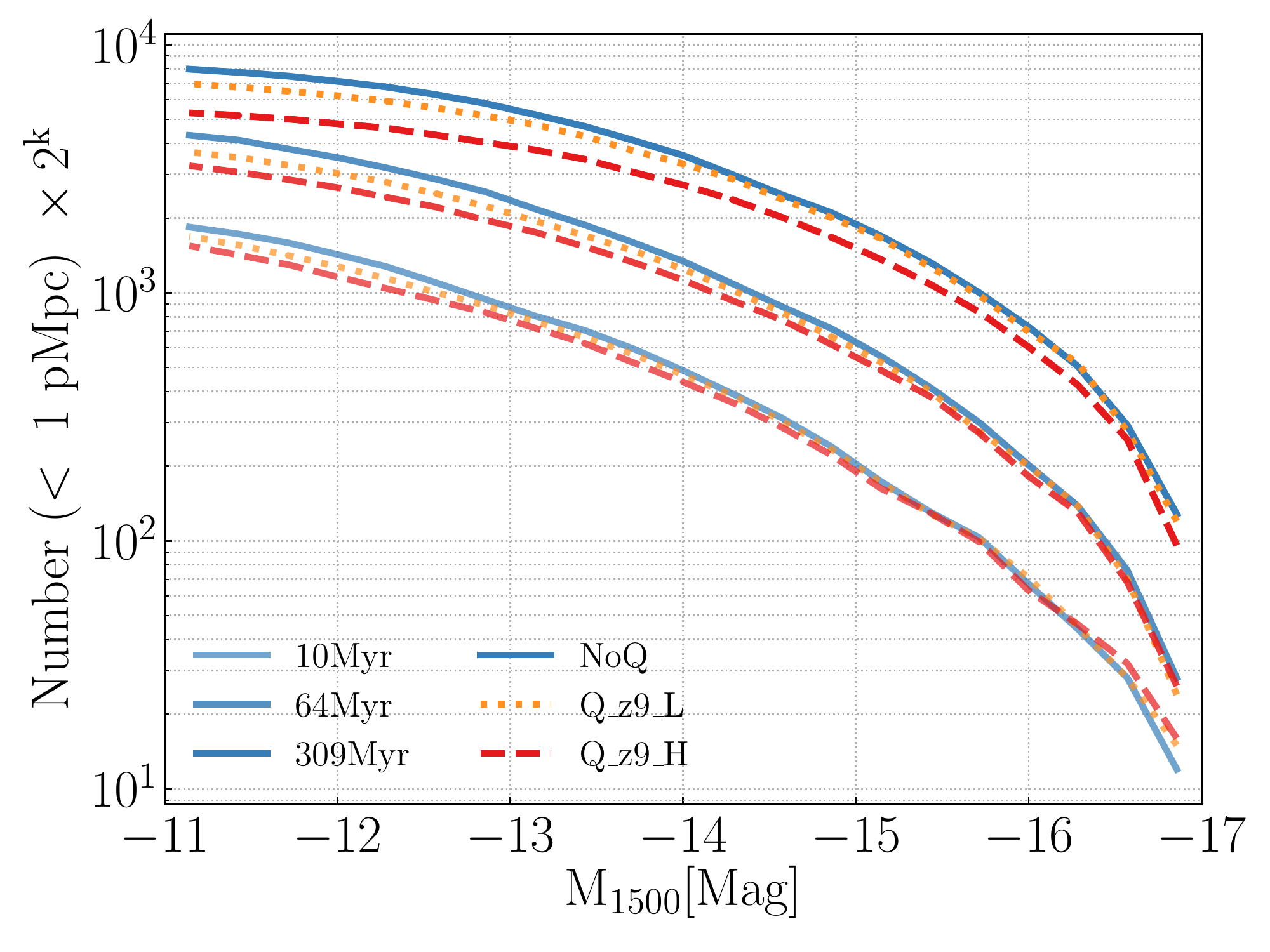}
\includegraphics[width=5.6cm]{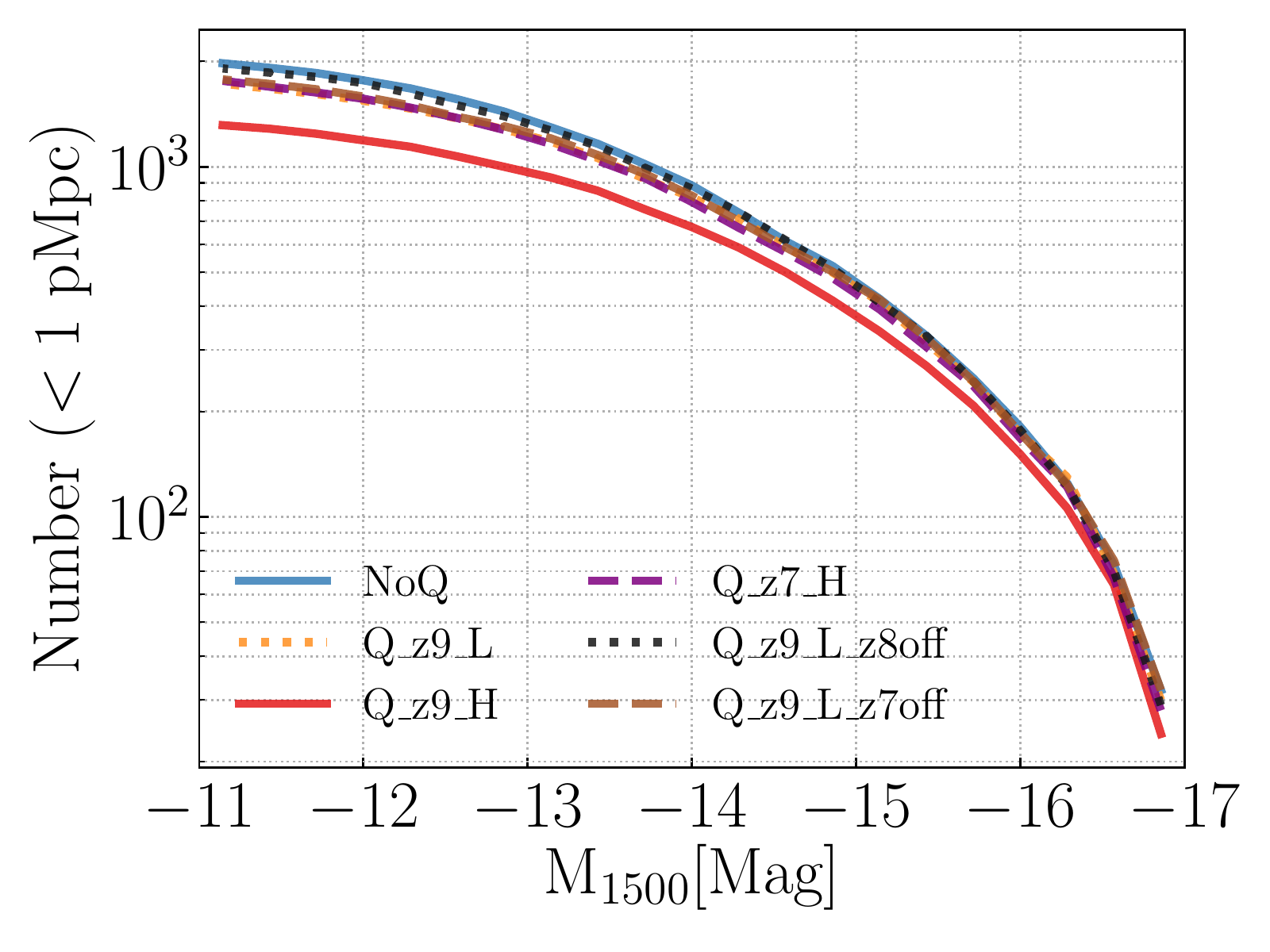}
\includegraphics[width=5.6cm]{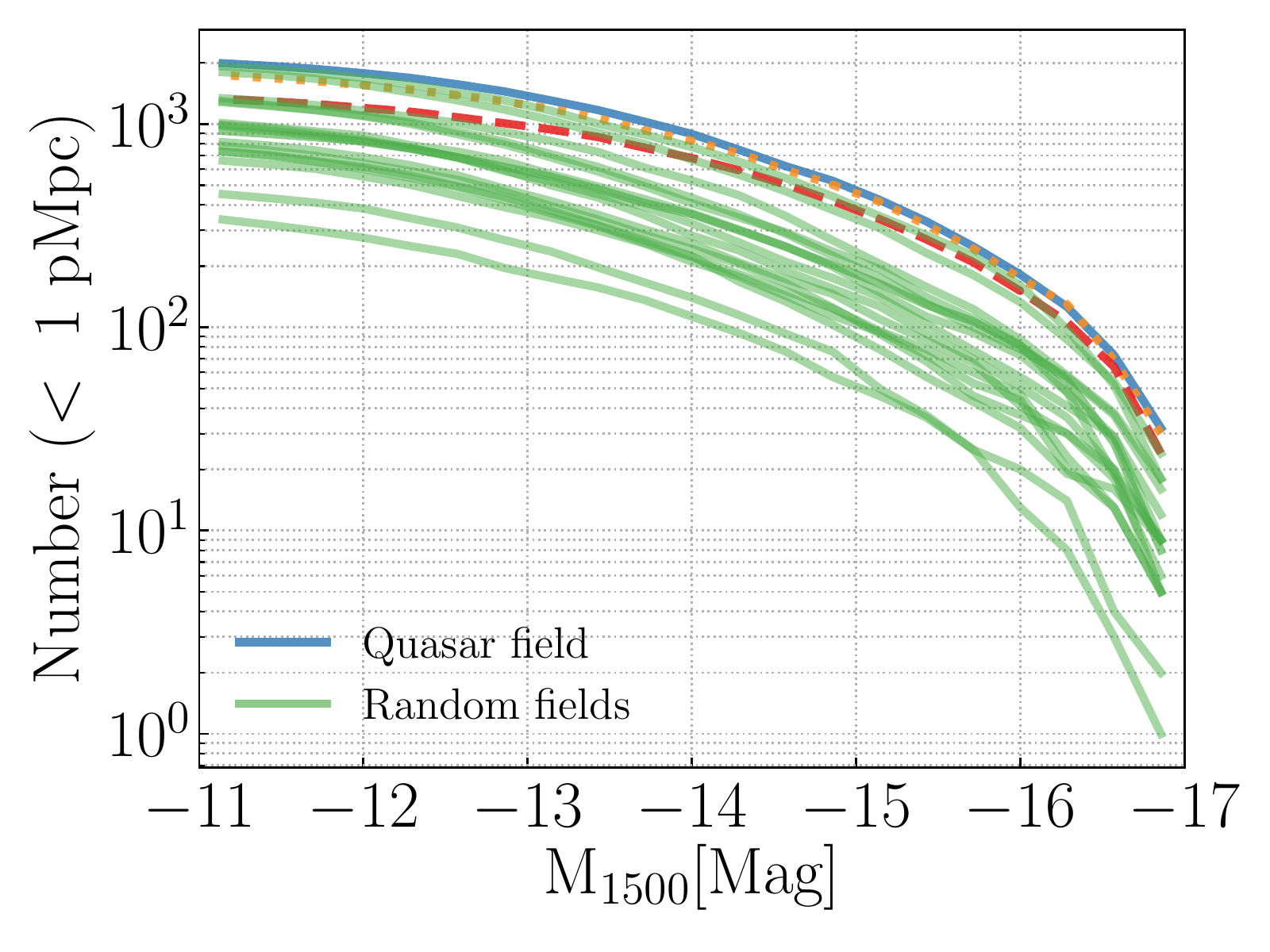}
\caption{Cumulative luminosity functions (CLFs) at wavelength 1500 \AA \ of galaxies  within a sphere of 1 pMpc centered on the quasar host. Left: CLFs at different quasar lifetimes in three simulations NoQ (blue), Q\_z9\_L (orange) and Q\_z9\_H (red). From bottom to top are for quasar lifetime of $10$ Myr, $64$ Myr and $309$ Myr. CLFs of $64$ Myr and $309$ Myr are manually multiplied by a factor of $2$ and $4$.  Middle: CLFs of all six simulations at $z=6.4$. Right: CLFs of NoQ, Q\_z9\_L, Q\_z9\_H at $z=6.4$ centered on the quasar, compared with CLFs (green) at the same time centered on $16$ different random locations in the simulation box. }\label{fig: LF}
\end{figure*}

\begin{figure}
\centering
\includegraphics[width=7cm]{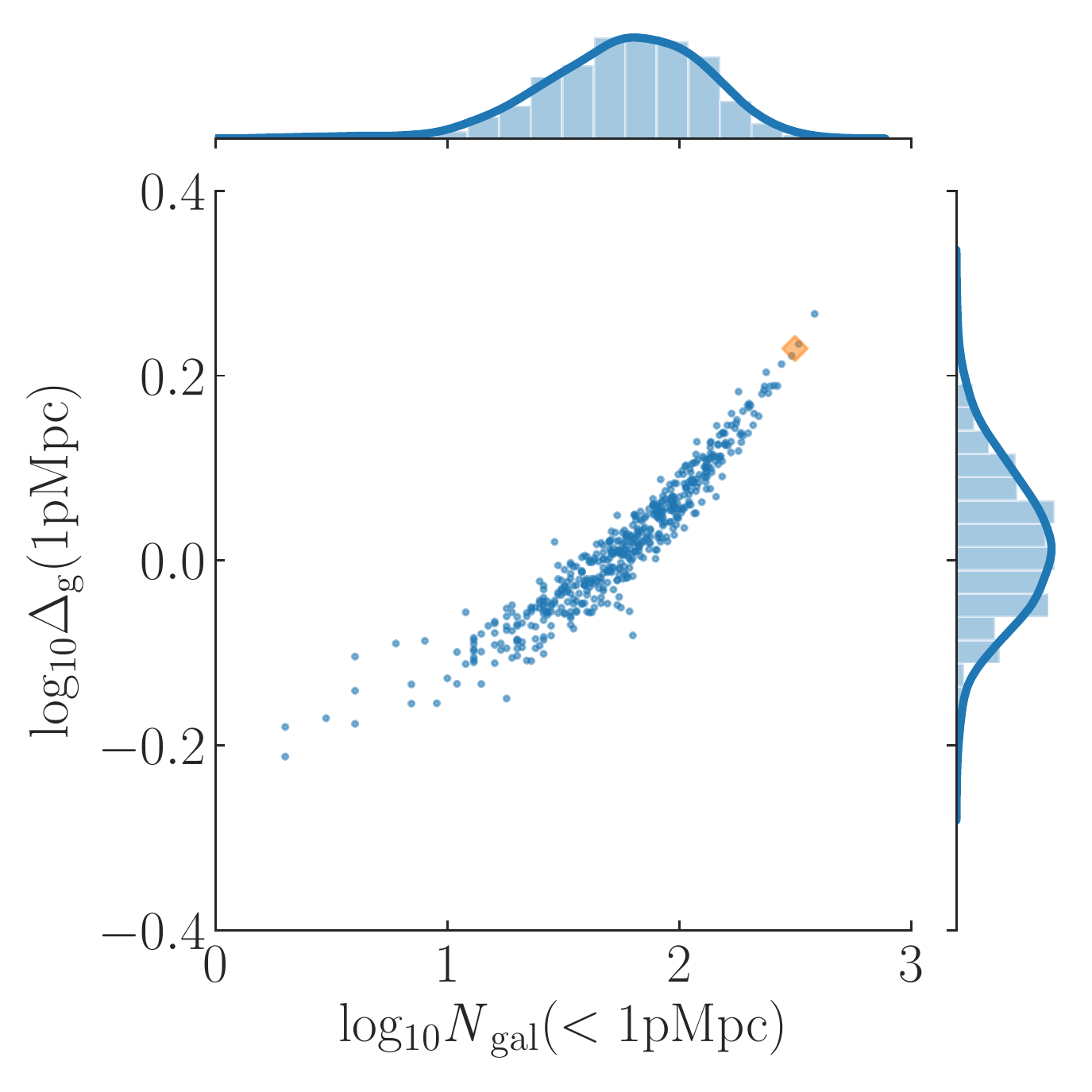}

\caption{$\log_{10} N_{\rm gal}$ and $\log_{10} \Delta_g$ distribution of 512 points uniformly sampling the simulation box NoQ, where $\log_{10} N_{\rm gal}$ is the number of galaxies with $\M1500<-16$ within a 1 pMpc sphere, and $\log_{10} \Delta_g$ is the gas mass overdensity smoothed by top-hat filter of the same size. The orange square is the position of the quasar host. The histogram on the top and right are for $\log_{10} N_{\rm gal}$ and $\log_{10} \Delta_g$ respectively.}\label{fig: Ng_od}
\end{figure}

\section{Observational Application}\label{sec:obs}
Luminosity function (LF) is a direct observable, and previous studies like \citet{ota18,kashikawa07} have measured the luminosity functions of LAEs. Due to the high cross-section of Ly$\alpha$ photons, to correctly model Ly$\alpha$ emission is very challenging. In contrast, theoretical modeling of the LF at some wavelength where the spectrum is continuous (e.g., at $1500$\AA) is straightforward. In the future, with the multiple wavebands, JWST will reliably measure the galaxy redshift and luminosity $L_{1500}$ and help us constrain the matter overdensity of the quasar fields.
Radiative feedback from quasars may complicate the procedure to use LFs to constrain the mass overdensity of quasar fields. However, as we see in Figure \ref{fig: QSO_SFH}, the impact on SFHs is small, especially for high mass halos. In this section, I examine the cumulative luminosity functions (CLFs) of galaxies in the simulations to study the degree of impact in detail.

In the left panel of Figure \ref{fig: LF}, I plot the CLFs in simulation NoQ, Q\_z9\_L and Q\_z9\_H at three different quasar lifetimes 10 Myr, 64 Myr and 309 Myr. I manually time the CLFs at 64 Myr by 2 and 309 Myr by 4 in order to put them in the same panel to compare. The most striking feature is that the degree of suppression is very small: even for the high luminosity quasar with 309 Myr on time, the suppression at $\M1500=-11$ is < 50\%. At $\M1500=-14$, where JWST will hopefully be able to measure the CLF, the suppression is only $\sim 25 \%$. This suppression increases with quasar lifetime but very slowly. For the low luminosity quasar case (orange dashed lines), the suppression on CLF at $\M1500=-11$ is $\lesssim 10$\%, even when the quasar is on for 309 Myr.  

In the middle panel of Figure \ref{fig: LF}, I plot the CLFs of all the six simulations at $z=6.4$. The suppression in simulation Q\_z9\_H is obviously the greatest because the quasar has the highest luminosity and the longest quasar lifetime. For the other four simulations, Q\_z9\_L\_z8off is almost exactly the same as the NoQ run, which is expected since the faint quasar is only on for $64$ Myr (see also Figure \ref{fig: turnoff}). The rest of the three simulations Q\_z9\_L, Q\_z9\_L\_z7off, and Q\_z7\_H are almost indistinguishable from each other. This means that 1) the short turn-off after a long quasar phase leaves negligible imprints on the CLF; 2) a quasar will leave similar imprints on the CLF, if its luminosity is ten times that of the lower luminosity quasar but its lifetime is an order of magnitude shorter.

In the right panel of Figure \ref{fig: LF}, I show how the CLFs in the quasar field differ from $16$ random fields at quasar lifetime $t_Q=309$ Myr . Random fields here are obtained by randomly casting  points in the simulation box NoQ and drawing spheres of $1$ pMpc centered on them. Compared to the field-to-field variation, the impact of quasar radiative feedback is very limited. As a result, we can use the CLF at 1500 \AA \ to constrain quasar field overdensity. In practice, since the CLF in random fields varies by an order of magnitude,  we need to observe many blank fields to robustly determine the average and variation of the CLF so that we can use them as references to determine the quasar field overdensity.

Since halos above $\M1500=-16$ are mostly not affected by quasar radiation, we can use these galaxies to probe the large scale environment where the quasar lives. I smooth the gas in the whole simulation box of NoQ by a spherical top-hat of radius $1$ pMpc and calculate the gas density contrast $\Delta_g$. Then I uniformly sample the box by $512$ points and for each point, I calculate the number of galaxies brighter than $\M1500=-16$ within the radius of $1 \rm pMpc$. In Figure \ref{fig: Ng_od} I show the distribution of this sample in blue dots. The orange diamond in this figure marks the position of the quasar host in my simulation suite, which lies on the very overdense tail. The two quantities $\log_{10} N_{\rm gal}$ and $\log_{10} \Delta_g$ shows a tight correlation, which means that the number of bright galaxies around quasar can be a good proxy of underlying matter density. The mean (standard deviation) of $\log_{10} N_{\rm gal}$ and $\log_{10} \Delta_g$ of this sample are $1.75(0.34)$ and $0.02(0.06)$, respectively. Because compared to the observable universe, my simulation box is too small which suffers from large cosmic variance, this may not represent the true distribution of the universe. With larger simulation boxes, we can model this distribution better. In the future, when more quasar fields and blank fields are observed, counting the number of bright galaxies ($\M1500<-16$) is a promising way to constrain the matter overdensity of quasar environment.

\begin{figure*}

 \centering
    \centering
     {\includegraphics[width=6cm]{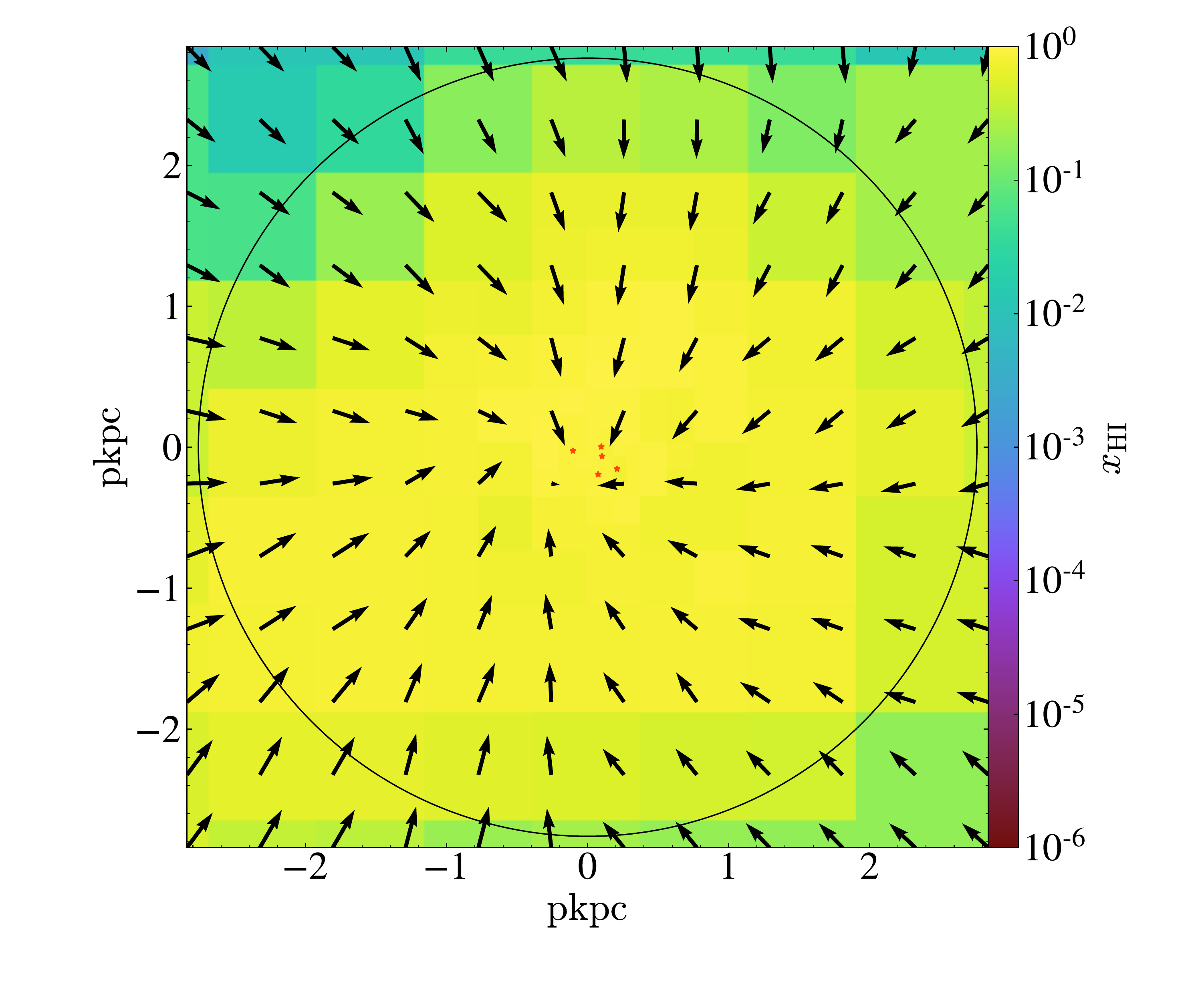}}\hspace{-0.4cm}
     {\includegraphics[width=6cm]{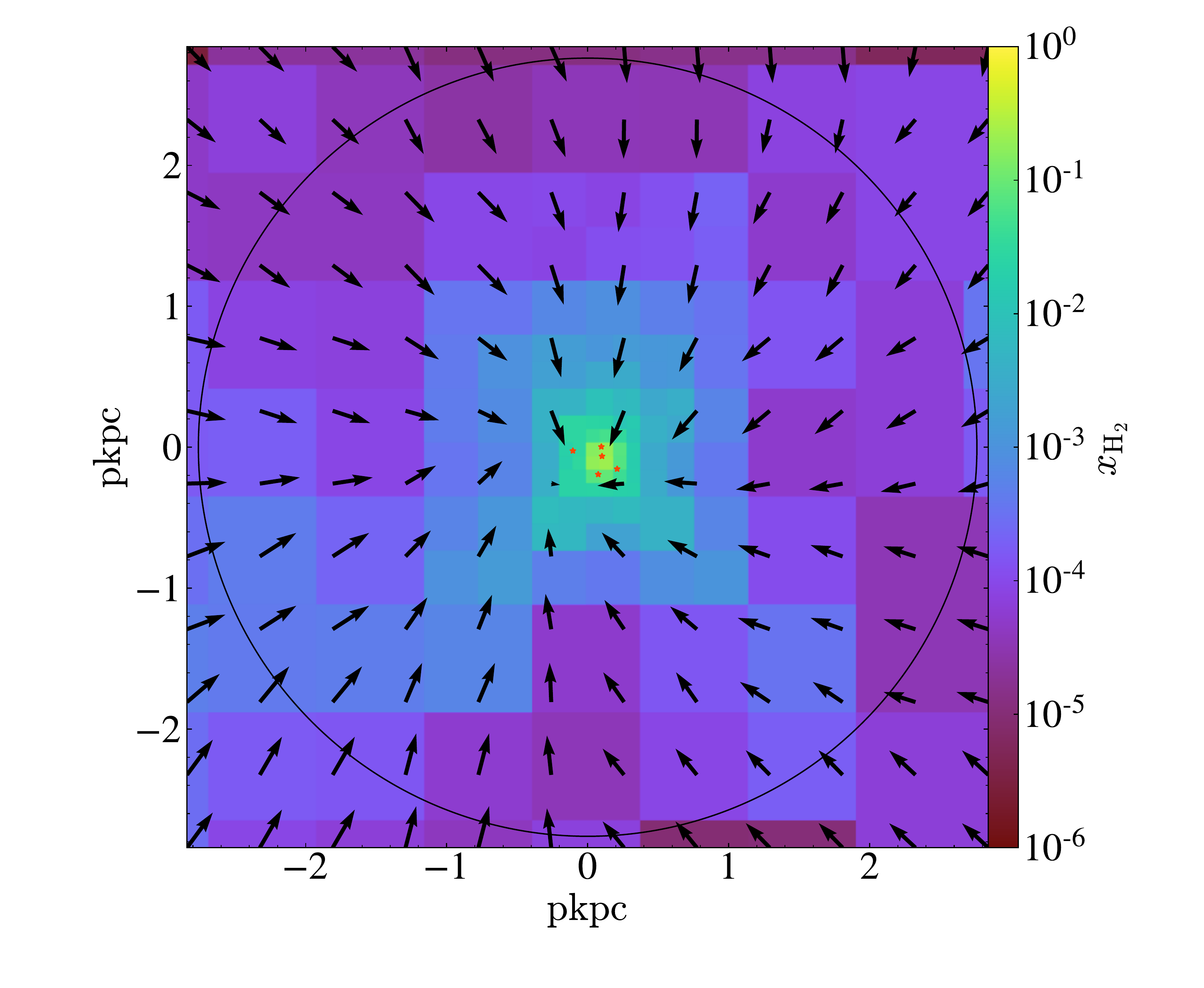}}\hspace{-0.4cm}
     {\includegraphics[width=6cm]{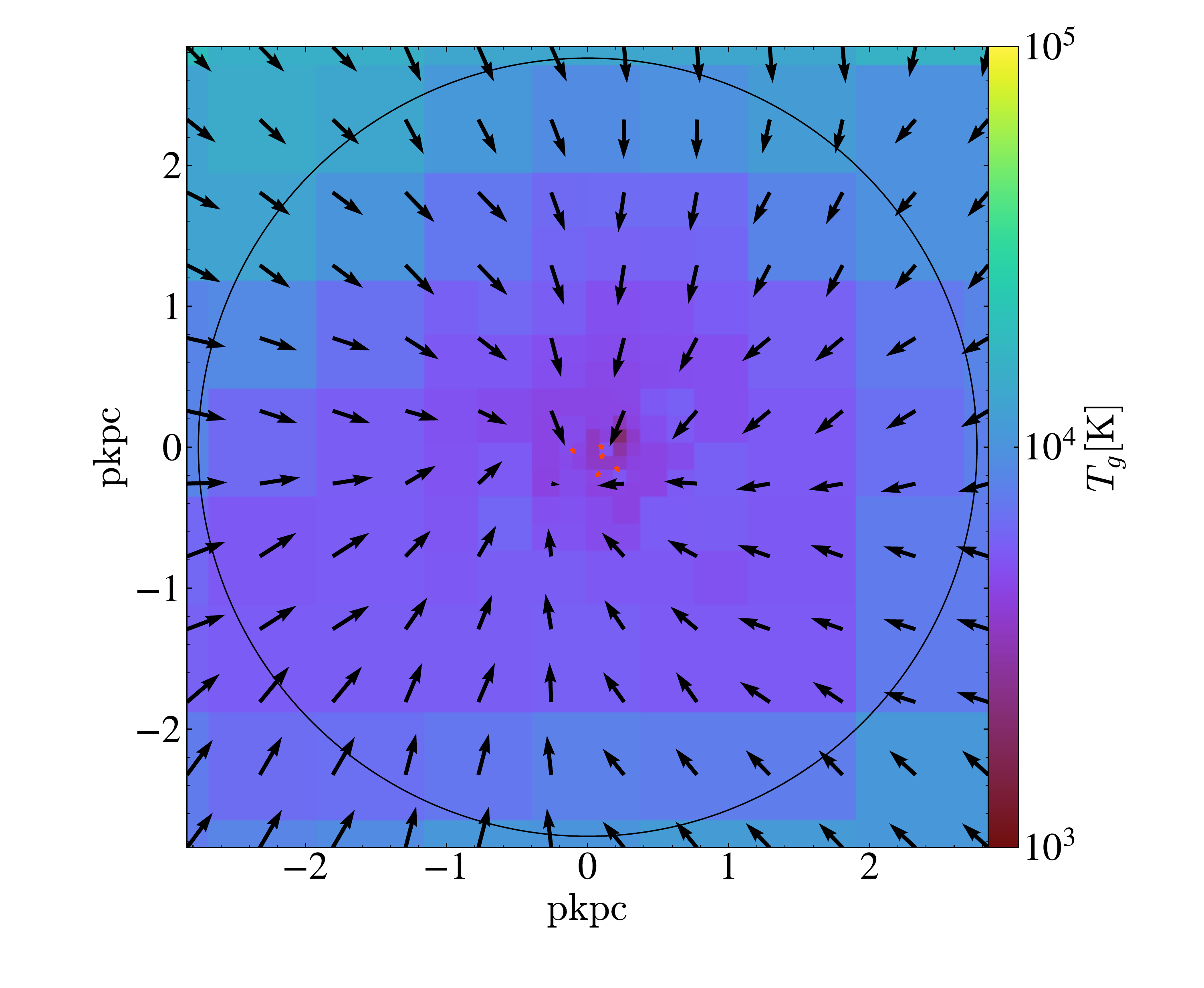}}
     
     {\includegraphics[width=6cm]{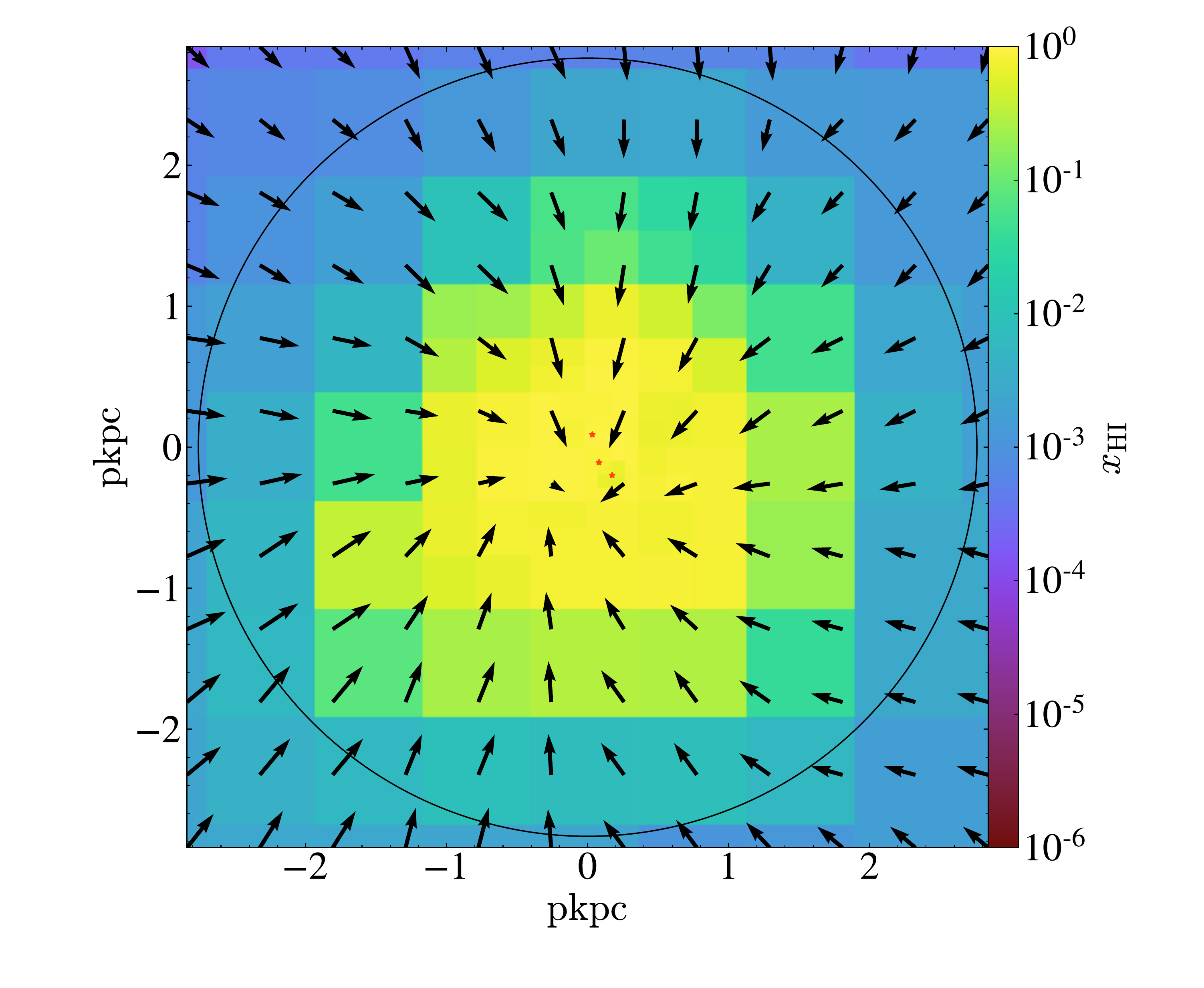}}\hspace{-0.4cm}
     {\includegraphics[width=6cm]{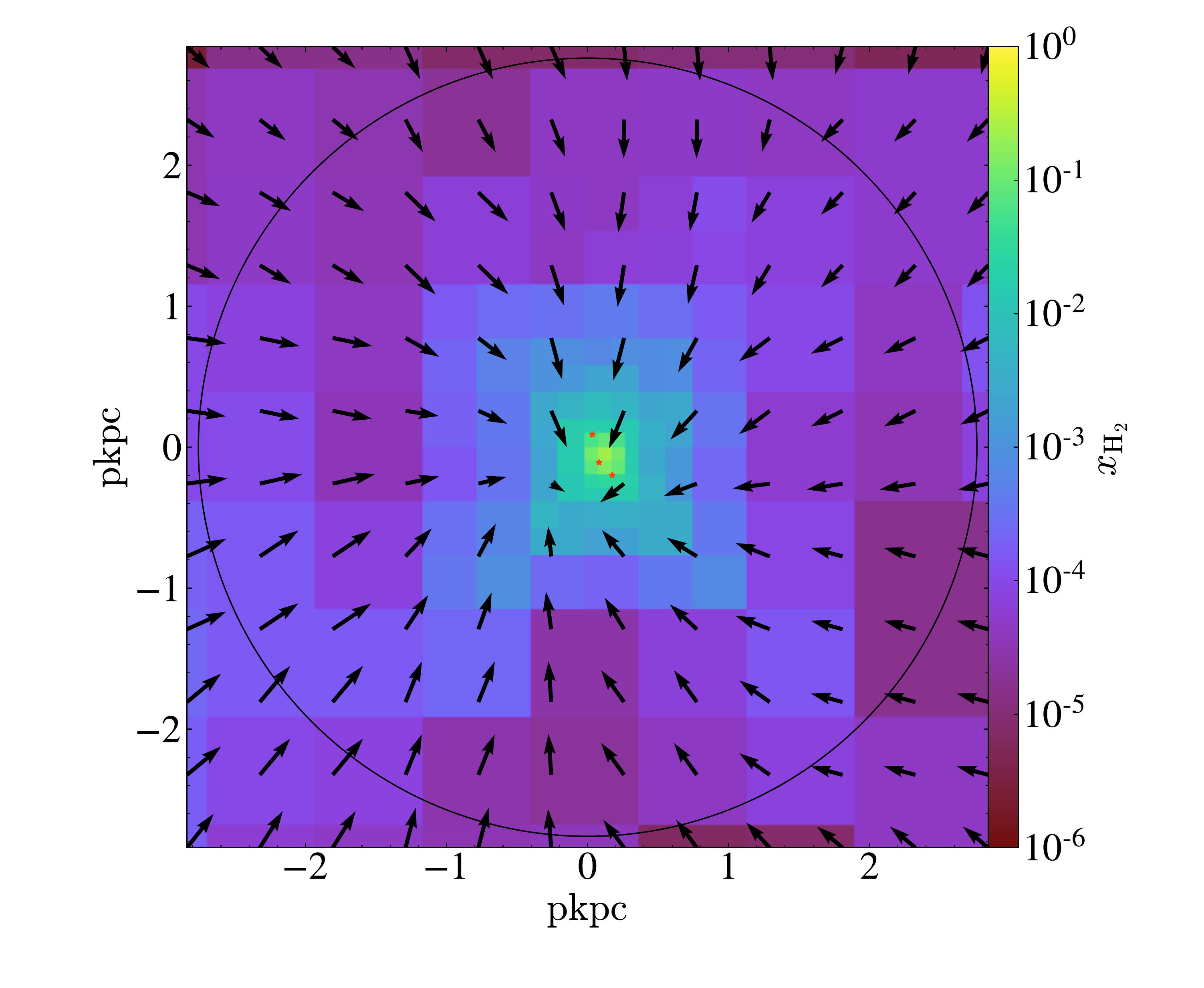}}\hspace{-0.4cm}
     {\includegraphics[width=6cm]{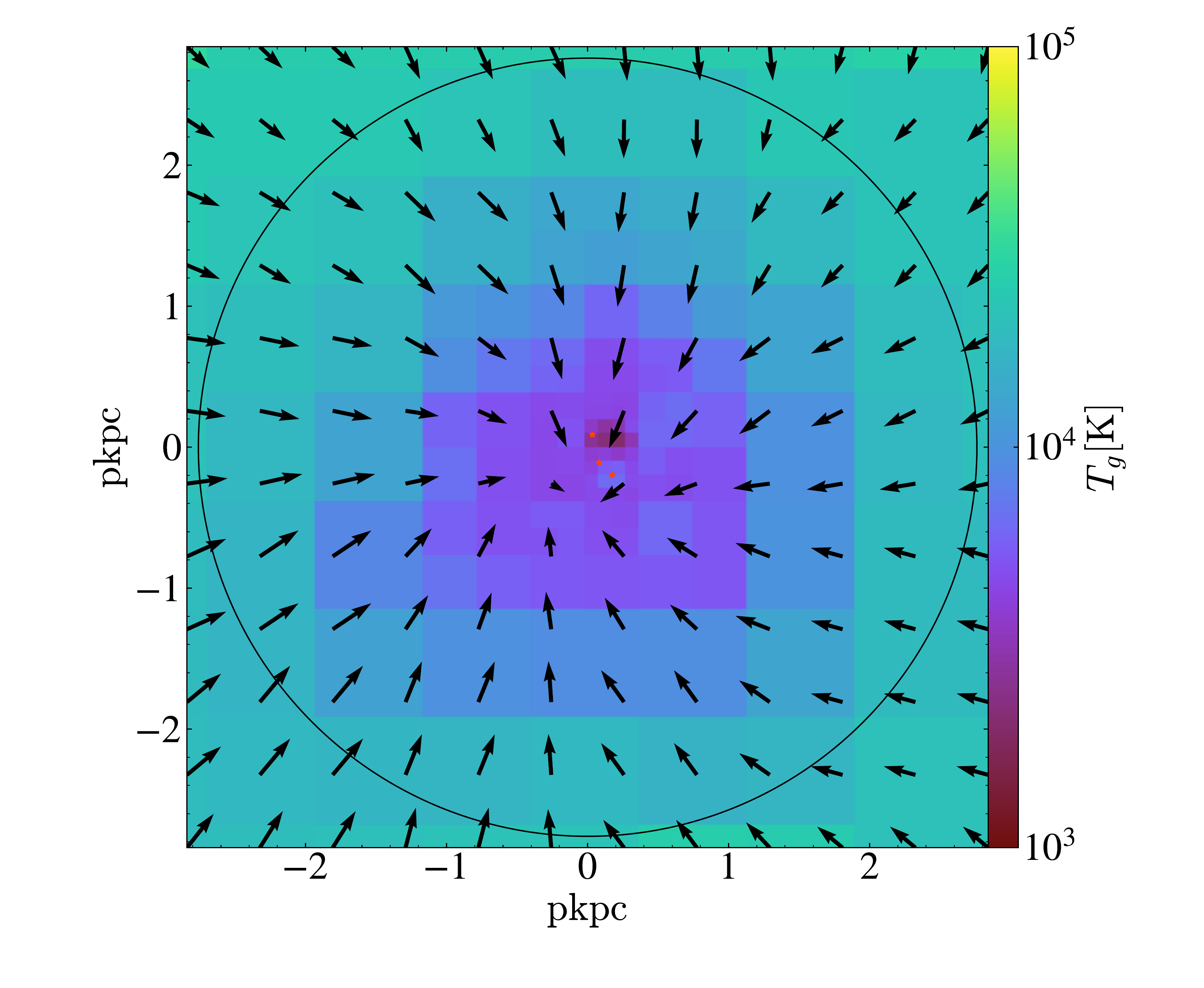}}

\caption{Upper panels: Neutral hydrogen fraction (left), molecular hydrogen fraction (middle) and gas temperature (right) maps a low mass halo Halo9677 in simulation NoQ. Black circle marks the halo radius at this time (z=8.2). Lower panels: Same as the upper panels but for simulation Q\_z9\_L. At this moment the quasar has been shining for $t_Q=64$ Myr. The overlaid arrows are gas velocities, and the red dots are young stars formed within 20 Myr.}\label{fig: visual_example_halo}
\end{figure*}

\section{Discussion}\label{sec:discussion}

\subsection{Subgrid model}
Because cosmological simulations typically have spacial resolution $\gtrsim 100$ pc, to model star formation, we need to adopt some subgrid recipes. Different subgrid recipes may result in different properties of galaxies. Therefore it is important to discuss some details of the subgrid model in this suite of simulation.

In my simulations, the stars form from molecular hydrogen, and the star formation rate is proportional to the $\rm H_2$ density in each cell:
    $$\rho_s = \frac{\rho_{\rm H_2}}{\tau}  $$
where $\tau$=1.5 Gyr is the depletion time. However, $\rm H_2$ chemistry is not explicitly modeled in the simulations, to a large degree because the spacial resolution 100 pc is too coarse to allow detailed $\rm H_2$ formation calculation.

One way to calculate the $\rm H_2$ mass is to use fitting formulas from higher resolution simulations. Here I use the subgrid model developed by \citet{gnedin11}. They have run a suite of zoom-in simulations to a Milky-Way sized halo. These simulations have a spatial resolution of 65 pc. In resolved scales, the radiation transfer, including the radiation transfer in Lyman-Werner bands, are done using the OTVET RT solver. In subgrid scales, Lyman-Werner radiation $I_{\rm LW}$ inside each simulation cell is modeled as 
$$I_{\rm LW} = S_{\rm H_2} \times S_{\rm dust} \times I^0_{\rm LW}$$
where $S_{\rm H_2}$ is the self-shielding factor of molecular hydrogen and $I^0_{\rm LW}$ is the ``free-space'' radiation field returned directly from the OTVET solver. $S_{\rm dust}$ is the shielding factor from dust,  

$$S_{\rm dust} = e^{ -D \sigma_0 (n_{\rm HI}+2n_{\rm H_2}) L_{\rm Sob} }$$
where $D$ is the dust-to-gas ratio, 
$\sigma_0 = 2\times 10^{-21} \rm cm$ 
is the dust cross-section, and 
$L_{\rm Sob}=\frac{\rho_g}{2|\nabla \rho_g|}$.
$S_{\rm H_2}$ is the self-shielding factor from molecular hydrogen. The calculation of $S_{\rm H_2}$ is complex since it is line absorption instead of continua absorption. This factor depends on gas density, internal velocity dispersion, molecular cloud size, radiation field, etc. \citet{gnedin14c} considered the supersonic turbulence and line overlap, parameterize the model including two free parameters sonic length and the clumping factor. These two free parameters are then fixed by calibrating against observations of Milky Way and Magellanic Clouds. They further ran a parameter grid of UV radiation intensity and dust-to-gas ratio and provided a fitting formula to calculate $\rm H_2$ fraction in each cell. This is the model used in my code.

To give some intuitive examples of how this model shapes the galaxies in my simulations, in Figure \ref{fig: visual_example_halo} I show the neutral hydrogen fraction (left panels), molecular hydrogen (middle panels) and temperature (right panels) of a small mass halo in simulations NoQ (upper panels) and Q\_z9\_L (lower panels) at $z=8.2$. At this moment the quasar has been on for 64 Myr. The overlaid arrows are gas velocities, and the red dots are young stars formed within 20 Myr. 
This halo is 0.43 pMpc away from the quasar host, and at this moment its halo mass is $5.6\times10^8\Msun$.  Comparing the upper and lower panels, we can find when the quasar turns on, although in the outskirts the hydrogen becomes highly ionized and temperature significantly increases, in the central region the gas remains cold and neutral. There is still molecular hydrogen in the center. Therefore the star formation does not stop completely despite the halo is small and close to the quasar.
Another thing worth to point out is that stars do not only form {in situ}. Stellar mass of a halo also grows when it accretes stars from surrounding subhalos. This can be another reason that halos do not stop increasing their stellar mass under quasar radiation.


\subsection{Comparison with previous study}
\citet{kashikawa07} used 1D spherical simulations \citep{kitayama00,kitayama01} to calculate the delay time of star formation of halos with masses ranging from $10^9 \Msun$ to $10^{11} \Msun$ under different UV intensities. They modeled the halo formation as a spherical cloud collapse, and defined the star formation time as 10\% of the baryonic mass has cooled below 1000K or density exceeds $10^6 ~\rm~ cm^{-3}$. In their Figure 8 they showed the delay time between runs without UV radiation and UV radiation turned on at z=8.3. They concluded that under UV intensity\footnote{$J_{\rm 21}=J / 10^{-21} ~\rm~ ergs cm^{-2} s^{_1} Hz^{-1} sr^{-1}$, where $J$ is the UV intensity at Lyman limit.} $J_{\rm 21}$=1, star formation is completely suppressed for halos smaller than $10^9 \Msun$. At higher $J_{21}=72$, star formation delayed for $100$ Myr for halos as massive as $10^{10} \Msun$.

It is hard to directly compare their results to this study, since in real scenarios halos are not isolated spheres and radiation intensity is not simply proportional to the inverse square of distance, due to the absorption of IGM. Nevertheless, here I choose halos in my simulation that are close to their input halo mass and UV intensity to compare. First, their halo collapse at $z=4.87$ while my simulations are run till $z=6.4$, so I use the average halo accretion histories from \citet{fakhouri10} to translate the halo mass at different redshifts. Halo masses $10^9 \Msun$ and $10^{10} \Msun$ at $z=4.87$ correspond to $6.5\times10^8 \Msun$ and $5.9\times10^{9} \Msun$ at $z=6.4$. As for UV intensity, on average, $J_{21}=1$ corresponds to 1 pMpc away from the quasar in run Q\_z9\_L and $J_{21}=72$ corresponds to 0.37 pMpc away from the quasar in run Q\_z9\_H. I check the SFHs of all halos of mass $6\times10^8 \Msun <M_h<7\times10^8 \Msun$ within a shell of distance $0.9 - 1.1$ pMpc from the quasar in run Q\_z9\_L, which results in 147 halos. These SFHs vary widely similar to the upper left panel of Figure \ref{fig: QSO_SFH}. Unlike predicted in \citet{kashikawa07}, they are not completely suppressed, although they all experience an apparent delay. The delay time is not a single value but varies from $\sim 10 $ Myr to infinity, with a typical value of $\sim 100$ Myr. To mimic the input value of $J_{21}=72$ and $M_h=10^{10} \Msun$ at $z=4.87$, I check the SFHs of halos of $5\times10^9\Msun<M_h<7\times10^9\Msun$ within $0.35 {\rm~ pMpc} < d < 0.4 {\rm~pMpc}$ from the quasar in run Q\_z9\_H. There are only four halos meet this requirement. Their SFHs flattens immediately after the quasar turns on, but star formation resumes again within less than $50$ Myr.

\begin{figure}[!htp]
     {\includegraphics[width=8.75cm]{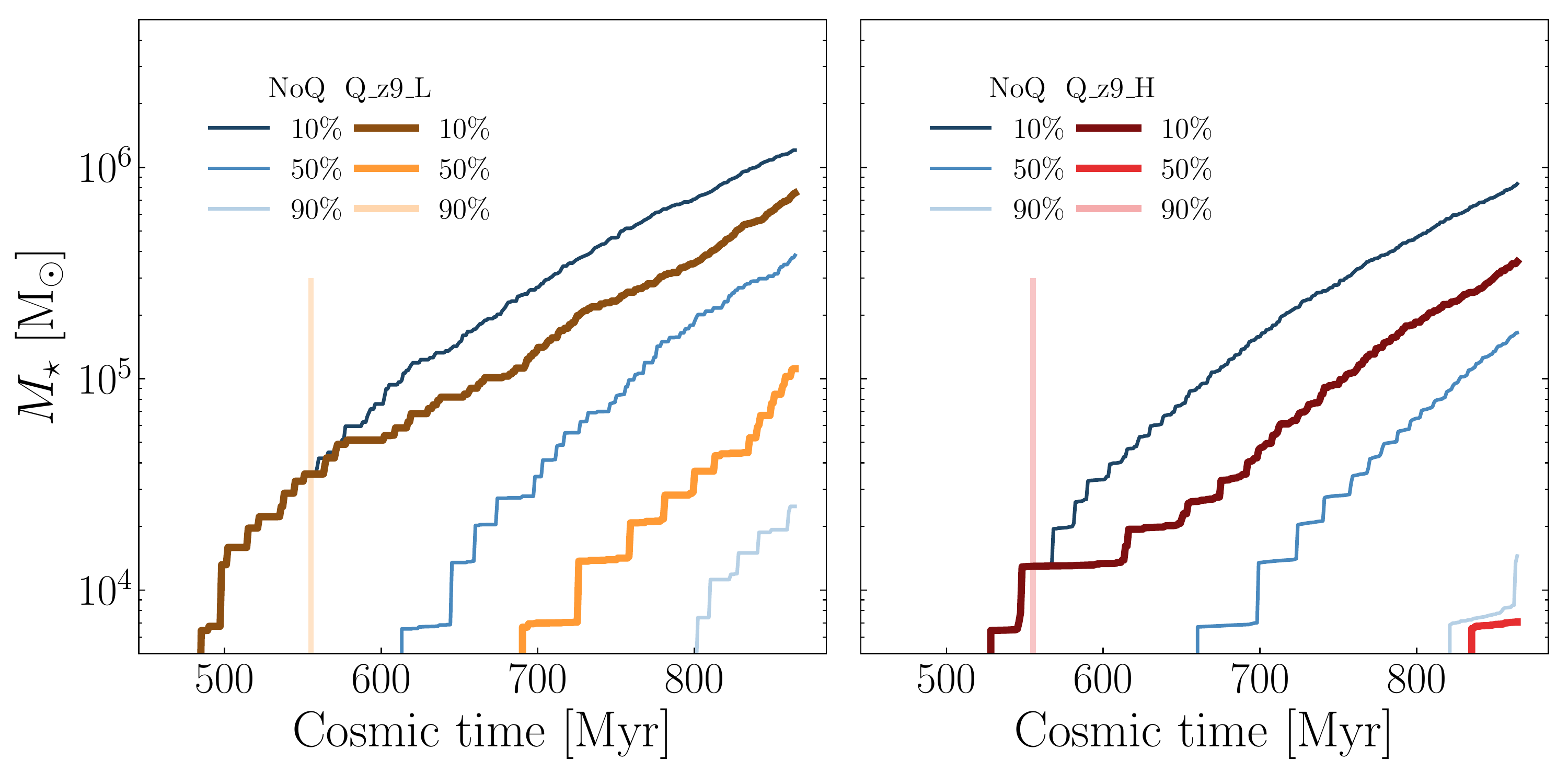}}
\caption{SFHs of halos with mass $5\times10^8 \Msun < M_h < 1\times10^9 \Msun$. Left: Halos are from a shell between $0.3 - 0.4$ pMpc from the quasar host in simulation NoQ (blue) and Q\_z9\_L (orange). Right: Halos are from a shell between $1 - 1.32$ pMpc from the quasar host in simulation NoQ and Q\_z9\_H (red). The halos in Q\_z9\_L run in the left panel are under similar quasar radiation intensity as those in Q\_z9\_H run in the right panel, since the quasar in Q\_z9\_H is $10$ times brighter but the halos are $3.3$ times further away.}\label{fig:SFH_dist_L_H}
\end{figure}

Solely comparing the star formation delay time, qualitatively, the results agree. However, one important thing we learn from this cosmological simulation suite is that even the halo mass and UV intensity is the same, the delay time in star formation varies because of the diversity in halo SFHs. To show this point more clearly, I select halos in the same mass bin $5\times10^8 \Msun < M_h < 1\times10^9 \Msun$ and under similar UV intensity in two quasar simulations. In Figure \ref{fig:SFH_dist_L_H}, plotted in the left panel in orange are SFHs of halos within the shell of $0.3 - 0.4$ pMpc from the quasar host in run Q\_z9\_L, while right in red are within the shell of $1 - 1.32$ pMpc in run Q\_z9\_H. I also plot the same halos in NoQ run in thin blue lines for comparison. We can see a difference in the ``delay time'' between these two. This is consistent with Figure \ref{fig:SFHdistance} and Figure \ref{fig:QSO_compare}: the halos closer to the quasar (left) form stars earlier, and thus become more ``resilient'' to quasar radiation.

\subsection{Limitation of simulation}

\section{Conclusion}\label{sec:conclusion}
In this study, I run a suite of 3D
 RT cosmological simulations to study galaxy evolution in quasar fields at $z=6.4$.
 The most massive halo at $z=9$ ($M_h=1.6 \times 10^{11} \Msun$) is chosen to host the quasar. By comparing halos with $M_h>5\times10^8 \Msun$ around the quasar in simulations with different quasar luminosities and lifetimes, I reach the following conclusions:
\begin{itemize}
    \item Low mass halos ($M_h<3\times10^9\Msun$) surrounding the massive quasar host halo have a large intrinsic variation of SFHs, even without quasar radiation. This variation also depends on the distance to the quasar.
    Within $\sim 1$ pMpc of the quasar host, low mass halos has intrinsically earlier star formation histories than those farther away.
    
    \item Quasar radiation suppresses SFR in low mass halos quickly through photo-dissociation of molecular hydrogen. Photo-heating contributes to the suppression after $\gtrsim 100$ Myr.
    \item Stellar mass is a good indicator of the degree of  suppression of star formation from quasar radiation. Quasar radiation does not have a significant impact on  halos within $1$ pMpc from the quasar which already have stellar mass above $1\times 10^5 \Msun$ at the moment the quasar just turns on (Figure \ref{fig:QSO_compare}). 
    \item Quasar radiative feedback suppresses the faint end of CLF by less than 25\% at $\M1500=-14$ even when the quasar is on for more than $300$ Myr. This difference in CLF caused by quasar radiation is far less than the field-to-field variation of CLF. 
    \item Quasar radiative feedback does not impact the bright end of the CLF. Therefore, the number of bright galaxies of quasar fields is a good indicator of the underlying mass overdensity.
\end{itemize}

This study is important for interpreting observations in quasar fields. It shows that quasars have a limited impact on the LF of galaxies surrounding them. Therefore, by comparing the luminosity functions between quasar fields and random fields we can effectively constrain the overdensity of the quasar environment. Future flagship JWST will map the galaxies around $z\gtrsim6$ quasars with better redshift and luminosity measurement, which will make major breakthroughs in learning the environments and properties of high-redshift quasars.

\acknowledgements
H.C. thanks Nick Gnedin for constructive feedback.
H.C. also thanks Jeff McMahon for support in manuscript preparation, through the Science Writing Practicum taught as part of U. Chicago's "Data Science in Energy and Environmental Research" NRT training program, NSF grant DGE-1735359.
This work was supported by a NASA ATP grant NNX17AK65G and NASA FINESST grant NNH19ZDA005K, and used resources of the Blue Waters sustained-petascale computing project, which is supported by the National Science Foundation (awards OCI-0725070 and ACI-1238993) and the state of Illinois. Blue Waters is a joint effort of the University of Illinois at Urbana-Champaign and its National Center for Supercomputing Applications. The testing phase of this project used the Midway cluster at the University of Chicago Research Computing Center.

{
\appendix
The current version of my code only allows one type of radiation source. 
Limited by this, in this study I use a galaxy-like spectrum for all sources including the quasar, which has no significant radiation about $4 \rm Ryd$. In this appendix, I briefly discuss the effect of a harder spectrum.

As discussed in the paper, the suppression in the early stage is mainly due to photo-dissociation of $\rm H_2$ by LW radiation. This is radiation in a narrow band $11.2 - 13.6 ~\rm~ eV$. Therefore, the spectrum shape above $13.6 ~\rm~ eV$ is unlikely to play a significant role in the early stage. 
To better show this point, I has run two simulations from $z=8.9$ to $z=8.2$ in which all radiation sources (including stars) are run with quasar-like spectra \citep{richards06}. 
Again, I run two simulations, with and without the quasar source, which I refer as NoQ\_Qspec and Q\_z9\_L\_Qspec, respectively. First, I calibrate the parameter depletion time $\tau_{\rm SF}$ and escape fraction below resolution limit $\epsilon_{\rm UV}$ so that the NoQ\_Qspec run has the similar star formation history and reionization history as the default NoQ simulation. Specifically, in both NoQ\_Qspec and Q\_z9\_L\_Qspec, $\tau_{\rm SF}$ and $\epsilon_{\rm UV}$ has been adjusted to $2.1 \rm Gyr$ and $0.3$, respectively, instead of $1.5 \rm Gyr$ and $0.15$ as in the original NoQ run.
In Q\_z9\_L\_Qspec, I turn on the quasar at the most massive halo at z=8.9, with the same ionizing radiation luminosity $L_Q=1\times10^{46} ~\rm~ erg/s$ as Q\_z9\_L. This translates to a LW radiation from the quasar in Q\_z9\_L\_Qspec $\sim 20\%$ of that in Q\_z9\_L. 
In the upper panels of Figure \ref{fig: appendix}, I plot the mean SFHs of halos in three halo mass (at $z=8.2$) bins $5\times10^8\Msun - 1\times10^9 \Msun$ (left), $1 \times 10^9 \Msun - 3\times10^9\Msun$ (middle), and $3\times10^9\Msun - 1\times10^{10}\Msun$ (right) from simulations NoQ\_Qspec, Q\_z9\_L\_Qspec, NoQ, Q\_z9\_L. The SFHs from NoQ\_Qspec and NoQ are similar by design. Comparing to a galaxy-like spectrum with the same $L_Q$, the suppression from a quasar-like spectrum is lower, which is expected because the lower radiation from LW band.

Simulations NoQ\_Qspec and Q\_z9\_L\_Qspec only run till $z=8.2$, thus can not address the effect from a harder spectra in timescale $\gtrsim 100 ~\rm~ Myr$. By ionizing H and He, quasar-like spectra can heat the gas to a higher temperature than galaxy-like spectra ($2.2\times10^4 ~\rm~K$ instead of $1.2 \times 10^4 ~\rm~K$, see also \citet{kakiichi17}). This may impact halos with slightly higher mass, but the bright end is still unlikely to be impacted significantly.  Harder photons like X-ray can also heat gas beyond the I-front of HII and HeIII, but they only bring up gas temperature by $\sim 3000 \rm K$.

\bibliography{ms}

\begin{figure*}
\centering
\includegraphics[width=\textwidth]{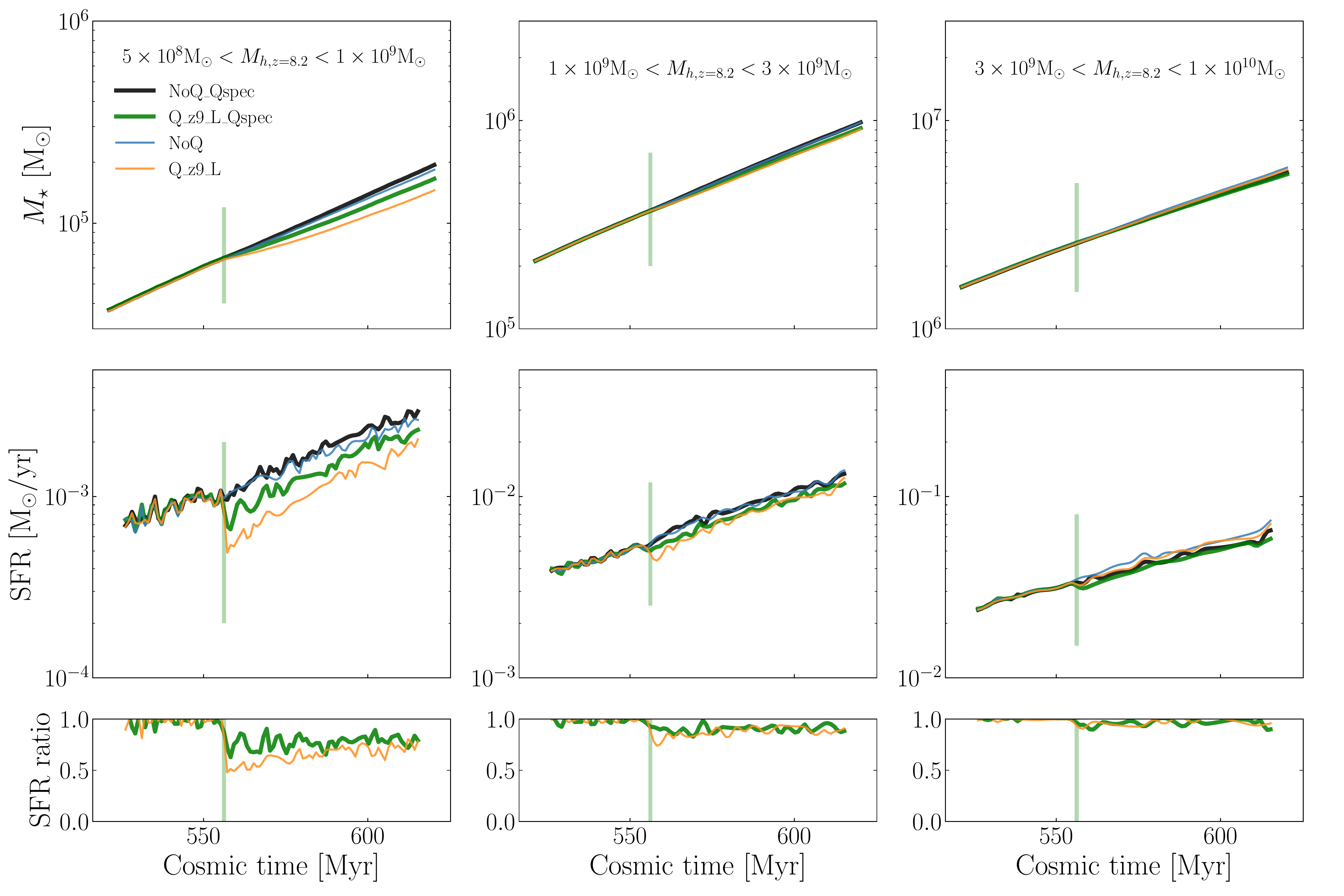}
\caption{Upper panels: mean SFHs of halos in three low mass bins (mass measured at $z=8.2$, or $t_Q=64 ~\rm~ Myr$). Halos are within a sphere of 1 pMpc from the quasar host. Black, green, blue and orange lines represent SFHs in simulations NoQ\_Qspec, Q\_z9\_L\_Qspec, NoQ, Q\_z9\_L, respectively. Middle panels: SFRs as a function of time in three low halo mass bins. Lower panels: star formation rate ratio of Q\_z9\_L\_Qspec/NoQ\_Qspec (green), Q\_z9\_L/NoQ (orange).
}\label{fig: appendix}
\end{figure*}

\end{document}